\documentclass[11pt, a4paper]{article}
\usepackage{mathtools,latexsym,color,graphicx,amsmath,amssymb,amsthm}

\def\reals{{\mathbb R}}
\def\eps{{\varepsilon}}

\def\S{{\cal S}}
\def\dist{{\sf dist}}
\def\dfds{{\sf dfds}}

\def\D{{\cal D}}

\def\S{{\cal S}}

\def\U{{\cal U}}

\def\marrow{\marginpar[\hfill$\longrightarrow$]{$\longleftarrow$}}
\def\micha#1{\textsc{(Micha says: }\marrow\textsf{#1})}
\def\matya#1{\textsc{(Matya says: }\marrow\textsf{#1})}

\newcommand{\old}[1]{{}}

\begin{document}
 
\newtheorem{theorem}{Theorem}[section]
\newtheorem{lemma}[theorem]{Lemma}
\newtheorem{corollary}[theorem]{Corollary}
\newtheorem{proposition}[theorem]{Proposition}

\begin{titlepage}

\title{Efficient algorithms for optimization problems involving semi-algebraic range searching}

\author{
Matthew J. Katz\thanks{%
Department of Computer Science, Ben Gurion University, Beer Sheva, Israel;
{\sf matya@cs.bgu.ac.il}
}
\and
Micha Sharir\thanks{%
School of Computer Science, Tel Aviv University, Tel Aviv Israel;
{\sf michas@tauex.tau.ac.il}
}
}

\date{}
 
\maketitle


\begin{abstract}
We present a general technique, based on parametric search with some twist, for solving
a variety of optimization problems on a set of semi-algebraic geometric objects of constant
complexity. The common feature of these problems
is that they involve a `growth parameter' $r$ and a semi-algebraic predicate $\Pi(o,o';r)$ of
constant complexity on pairs of input objects, which depends on $r$ and is monotone in $r$, meaning
that if $\Pi(o,o';r_1)$ is true then $\Pi(o,o';r_2)$ is true for any $r_2 > r_1$. One then defines
a graph $G(r)$ whose edges are all the pairs $(o,o')$ for which $\Pi(o,o';r)$ is true, and
seeks the smallest value of $r$ for which some graph-monotone property holds for $G(r)$.

Problems that fit into this context include (i) the reverse shortest path problem in unit-disk 
graphs, recently studied by Wang and Zhao~\cite{WZ}, (ii) the same problem for weighted unit-disk graphs,
with a decision procedure recently provided by Wang and Xue~\cite{WangX20}, 
(iii) extensions of these problems to three and higher dimensions, 
(iv) the discrete Fr\'echet distance with one-sided shortcuts in higher dimensions, 
extending the study by Ben Avraham et al.~\cite{BFKKS}, 
(v) perfect matchings in intersection graphs: given, e.g., a set of fat ellipses of roughly 
the same size, find the smallest value $r$ such that if we expand each of the ellipses by $r$ 
(either additively or multiplicatively), the resulting intersection graph contains a perfect 
matching, (vi) generalized distance selection problems: given, e.g., a set of disjoint segments, 
find the $k$'th smallest distance among the pairwise distances determined by the segments, 
for a given (sufficiently small but superlinear) parameter $k$ and an appropriate definition 
of distance between segments, and (vii) the maximum-height independent towers
problem, in which we want to erect vertical towers of maximum height over a 1.5-dimensional
terrain so that no pair of tower tips are mutually visible.

We obtain significantly improved solutions for problems (i), (ii) and (vi),
and new efficient solutions to the other problems, which do not appear to have been studied earlier. 

In general, our technique, when applicable, produces solutions that are significantly more efficient 
than those obtained by parametric search (or one of the alternative techniques), and unlike parametric 
search it does not require any parallelism. 
 
\end{abstract}

\end{titlepage}


\section{Introduction} \label{sec:int}

In this work we generalize, in a very broad and comprehensive manner, a technique that 
was originally developed in Ben Avraham et al.~\cite{BFKKS} for solving the discrete Fr\'echet distance
problem with one-sided shortcuts. We show that the high-level idea behind the technique is 
sufficiently versatile, so that, with suitable and rather nontrivial enhancements, it can 
be applied to many other optimization problems that have the underlying structure described 
in the abstract. Specifically, the input to these problems is a set $S$ of $n$ semi-algebraic
geometric objects of constant complexity, in the plane or in higher dimensions,
and a semi-algebraic predicate $\Pi(o,o';r)$ of constant complexity on pairs of input objects,
that also depends on a `growth parameter' $r>0$. The predicate $\Pi$ is assumed to be monotone 
in $r$, meaning that if $\Pi(o,o';r_1)$ is true for some pair $o,o'\in S$ then $\Pi(o,o';r_2)$ 
is true for any $r_2 > r_1$. Thus each pair $o,o'\in S$ has an associated \emph{critical value}
$r_{o,o'}$, which is $\min \{ r \mid \Pi(o,o';r) \text{ is true} \}$ (in all our applications
the minimum is attained). Define a `proximity graph' $G(r)$ on $S$, whose edges are all the 
pairs $(o,o')$ for which $\Pi(o,o';r)$ is true. The optimization problem is to find
the smallest value $r^*$ of $r$ for which some graph-monotone property holds for $G(r^*)$.

As an example that clarifies these concepts consider the 
\emph{reverse shortest path problem for unit-disk graphs}, 
recently studied by Wang and Zhao~\cite{WZ}. In this problem we are 
given a set $P$ of $n$ points in the plane, two points $s,t\in P$, and
an integer parameter $k$, and we want to find the smallest value of $r$
for which there is a path of at most $k$ edges between $s$ and $t$ in
$G(r)$, which, in this special case, is the graph over $P$, whose edges
are all the pairs $(u,v)$ of points of $P$ at distance at most $r$.
The predicate $\Pi(u,v;r)$ is simply $\|u-v\|\le r$, where $\|\cdot\|$ 
denotes the Euclidean distance; clearly $\Pi$ is monotone in $r$, and the
existence of a path of length $k$ in $G(r)$ is a graph-monotone property.

Wang and Zhao present a solution to this reverse shortest path problem 
that runs in $O^*(n^{5/4})$ time (where the $O^*(\cdot)$ notation 
hides subpolynomial factors),
in which the decision procedure takes linear time. We improve their
result and obtain an algorithm that runs in randomized expected $O^*(n^{6/5})$ time.

We will shortly list many additional problems that fall into our context and the
improved or new performance bounds that our technique yields.

Roughly speaking, one can think of our technique as a variant of parametric search,
which aims to find the optimum value $r^*$ of $r$, on which the associated decision 
procedure is based. We assume that this procedure, which determines whether its input 
parameter $r$ is larger than, smaller than, or equal to $r^*$, is efficient, but we 
do not have an efficient way to parallelize it, which is required by standard parametric 
search to make the optimization procedure efficient. That is, if we could parallelize 
the decision procedure, say with polylogarithmic depth, we could apply standard 
parametric search and solve the optimization problem in time that is roughly the 
same as the cost of the decision procedure, up to a polylogarithmic factor.
There are variants of parametric search, e.g., an expander-based approach~\cite{KS},
or techniques based on random sampling~\cite{Chan01}, where parallelism is not required. 
Nevertheless, standard implementations of these techniques also do not seem to achieve the
improved bounds that we obtain here. 
For example, in a sampling-based approach, we want to take a random sample of
a few critical values of $r$, and run binary search through these values
to narrow the range that contains the optimum $r^*$. This is difficult to do when
we only have a subrange of values from which we want to sample.

Standard parametric search simulates the execution of the decision procedure
at (the unknown) $r^*$, so that it determines the outcome of each comparison made by the 
procedure at $r^*$, by comparing $r^*$ with the $O(1)$ critical values of the 
comparison, at which it changes its sign, applying the (unsimulated) decision
procedure itself at each (or at some) of these critical values to resolve the 
comparison. Instead, our variant employs 
a different approach, and simulates the decision procedure without resolving 
each of its comparisons right away. This results in a \emph{bifurcation tree}, 
in which all possible outcomes ($r > r^*$, $r < r^*$, $r=r^*$) are explored. 
Only at certain steps during the execution we stop the construction of the tree, 
resolve all the comparisons that the algorithm has encountered, and are represented 
in the tree, follow the unique path in the tree that results from these comparisons
(which represents a portion of the execution of the decision procedure at $r=r^*$),
and then repeat this process, until the entire simulation of the decision procedure is completed.

In general, this technique would be too expensive, but we ensure its efficiency 
by applying it within a range $I = (\alpha,\beta]$ of values of $r$ that contains 
$r^*$ and only a prespecified small number $L$ of additional critical values. 
This will ensure that in most of the comparisons that the bifurcation tree 
encounters there will be no bifurcation because the relevant value of $r$
will lie outside $I$, so its relation to $r^*$ is known. 
The `interval-shrinking' stage, which precedes the bifurcation-based simulation
of the decision procedure, is performed using
a fairly extensive generalization of the technique in \cite{BFKKS}. That technique 
was designed in \cite{BFKKS} for the special case where the critical values are 
distances between points in the plane, but we show that it can be extended to any 
kind of criticalities, as long as the critical values can be specified by a 
constant-complexity semi-algebraic predicate, which is the general situation 
that we assume here. This expansion of the context in which this technique can 
be applied is one of the main contributions of our paper. 

The running time of the original procedure in \cite{BFKKS}, for distances between 
points in the plane, is $O^*(n^{4/3}/L^{1/3})$. In the more general setup considered 
here, the performance depends on the number of degrees of freedom of the input objects 
and on the cost of the decision procedure. For example, for distances between points 
in $\reals^3$ and a near-linear decision procedure, the performance is 
$O^*(n^{3/2}/L^{1/2})$. See below for full details.

Combining the interval-shrinking step with the bifurcation-tree procedure, a 
careful choice of $L$ results in an efficient algorithm, whose running time 
depends on the number of degrees of freedom of the input objects and on the
complexity of the decision procedure. For example, for distances between
points in the plane and for a near-linear-time decision procedure, the 
algorithm runs in $O^*(n^{6/5})$ time.


We demonstrate the applicability of the resulting technique to numerous
optimization problems that we now proceed to list.

\medskip
\noindent{\bf (i) The reverse shortest path problem for unit-disk graphs.}
This already mentioned problem was recently studied by 
Wang and Zhao~\cite{WZ}. Recall that in this problem we are given a 
set $P$ of $n$ points in the plane, two points $s,t\in P$, and an 
integer parameter $k$, and we want to find the smallest value of $r$
for which there is a path of at most $k$ edges between $s$ and $t$ in
$G(r)$, which is the graph over $P$, whose edges
are all the pairs $(u,v)$ of points of $P$ at distance at most $r$.
Wang and Zhao present a solution that runs in $O^*(n^{5/4})$ time,
in which the decision procedure takes linear time. We improve their
result and obtain an algorithm that runs in $O^*(n^{6/5})$ randomized expected time.

\medskip
\noindent{\bf (ii) The reverse shortest path problem for weighted unit-disk graphs.}
We then consider the weighted variant of this problem, in which we
want to find the minimum value $r^*$ such that $G(r^*)$, defined as above,
contains a path from $s$ to $t$ of length at most $w$, where each edge
of $G(r^*)$ has a weight equal to its Euclidean length, and the
length of the path is the sum of the weights of its edges. The decision
problem for this task, in which we specify $r$ and seek the shortest path
in $G(r)$ from $s$ to $t$, has been considered by Wang and Xue~\cite{WangX20},
who gave an algorithm that takes $O(n\log^2n)$ time. Applying our
technique and using this decision procedure, we solve the weighted variant of
the reverse shortest path problem for unit-disk graphs in $O^*(n^{6/5})$ randomized expected time.
Previously, Wang and Zhao~\cite{WZ} observed that this variant can be solved 
in $O(n^{4/3}\log^2 n)$ time, and mentioned the question of whether the 
$O(n^{4/3})$-time barrier can be broken as an interesting open problem.  

\medskip
\noindent{\bf (iii) Extensions of the reverse shortest path problems to three and higher dimensions.} 
We then consider the reverse shortest-path problems from \cite{WZ} and \cite{WangX20}, 
in three and higher dimensions. To the best of our knowledge, distance 
selection, on which the method is based for the case of distances between points,\footnote{%
  More precisely, the technique is based on a time-improving modification 
  of the procedure for distance selection, namely the interval-shrinking
  procedure mentioned earlier.}
has not been explicitly studied in higher dimensions. This is a simple example
of the way in which the technique of \cite{BFKKS} can be extended. More involved
examples will be listed shortly.

The extension of each of the two reverse shortest-path problems (unweighted and weighted)
to higher dimensions is interesting also because the decision procedure turns out 
to be more expensive and takes superlinear time, which is nonetheless faster than 
the cost of distance selection, on which the interval-shrinking part is based.
As we argue, this suffices for obtaining an improved solution, whose
running time lies `in between' the two costs.

We solve both problems in three dimensions in $O^*(n^{17/12})$ randomized expected time.
We sketch an extension of this result to any $d$ dimensions for the unweighted case, which runs in
${\displaystyle
O^*\left(n^{\frac{(6d-4)\lfloor (d+1)/2\rfloor + 2d} {(3d-1)\left(\lfloor (d+1)/2\rfloor + 1 \right)} }\right) }$
randomized expected time, which is indeed faster than the cost $O^*(n^{2d/(d+1)})$ of distance selection.

\medskip
\noindent{\bf (iv) The discrete Fr\'echet distance with shortcuts in higher dimensions.} 
We then consider the Fr\'echet problem from \cite{BFKKS} in $\reals^d$, for any $d\ge 3$.
In contrast with the reverse shortest path problem, the Fr\'echet distance problem has a 
decision procedure which is linear in any dimension. The cost of the distance selection
procedure keeps growing with the dimension. The cost of the optimization 
procedure keeps growing too, but is always significantly smaller than the
cost of distance selection. Concretely, our algorithm runs in 
$O^*(n^{(4d-2)/(3d-1)})$ randomized expected time.

\medskip
\noindent{\bf (v) Perfect matchings in intersection and proximity graphs.}
We next consider applications of a different kind, involving intersection graphs
of geometric objects. In these problems the critical values, determined by pairs 
of objects, are more involved than inter-point distances, and the underlying monotone 
graph property is the existence of a perfect matching in the graph.

Let $\U$ be a set of $n=2m$ disks in the plane with radii in the 
interval $[1,\Psi]$ for some constant parameter $\Psi$. The problem is to
find the smallest value $r^*$ so that if we blow up each disk about its center
by $r^*$, either additively or multiplicatively, the intersection graph of the
expanded disks has a perfect matching.

Bonnet et al.~\cite{BCM} present an algorithm for computing a matching in the 
intersection graph $G$ of $\U$ that, with high probability, is a maximum 
matching. The algorithm runs in 
$O(\Psi^6 n \log^{11} n + \Psi^{12\omega}n^{\omega/2}) = O(n^{\omega/2})$
expected time, where $\omega\approx 2.37286$ is the exponent of matrix multiplication.

Let $\U_r$ denote the set of the disks of $\U$, each expanded by $r$. That is,
we either add $r$ to its radius or multiply it by $r$. We apply the procedure
of \cite{BCM} to the intersection graph $G(r)$ of $\U_r$ as a decision procedure
for the search of the optimum value $r^*$. We observe that the requirement that 
the radii are between 1 and $\Psi$ does not cause a problem. Indeed, when the
blow-up of the disk radii is multiplicative, the ratio between the largest and 
smallest radii does not change and remains $\Psi$. When the blow-up is additive, 
the ratio only goes down, assuming that $r^* > 0$ (that is, the intersection
graph of the original disks does not contain a perfect matching).

We show that one can compute the smallest $r^*$ such that $G(r^*)$ contains a 
perfect matching, in $O^*(n^{3/4+\omega/4}) \approx O(n^{1.3432})$ randomized 
expected time. To appreciate this bound, one should compare it with the appropriate 
bound for \emph{critical value selection}, which is $O^*(n^{3/2})$; see a remark 
following Theorem~\ref{th:shrink}. 

\old{
\micha{Why does this para hold? Are you relying on Lemma 13 of \cite{BCM}? 
Is it clear how to implement the dynamic data structure that they need?} \matya{This should be checked}
Actually, the $O(n^{\omega/2})$ bound mentioned above holds for any set of \emph{fat} objects 
of similar sizes, provided that one can perform some basic operations on these
objects efficiently. For example, suppose that $\U$ is a set of $n=2m$ ellipses 
in the plane, such that for each ellipse in $\U$, the ratio between its major 
and minor axes is at most some constant parameter $\Psi_1$, and the length of its major
axis is in $[1,\Psi_2]$, for some constant parameter $\Psi_2$. Then
one can compute the smallest $r^*$ such that $G(r^*)$ contains a perfect matching, 
in $O^*(n^{(5+2\omega)/7}) \approx O^*(n^{1.3922})$
expected time. If the ratio between the main and secondary axes 
of the ellipses is some constant, e.g., 2, then
we can do it in $O^*(n^{(8+3\omega)/11}) \approx O^*(n^{1.3744})$ expected time.
}

\old{
\matya{This is not interesting since in any fixed dim $d$, there exists a constant $c(d)$, such that the $c(d)$RNG of the underlying point set contains a bottleneck matching.}

Let $\U$ be a set of $n=2m$ unit balls in $\reals^d$. Bonnet et al~\cite{BCM} 
present a randomized algorithm that computes a maximum matching in the intersection 
graph $G$ of these balls in $O(n^{3/2})$ time, for $d =3,4$, and in 
$O^*\left(n^{\frac{2\lceil d/2 \rceil}{1+\lceil d/2 \rceil}}\right)$ time, 
for $d \ge 5$. (The sparsification step in their algorithm, which replaces
the intersection graph by a subgraph with a linear number of edges, takes $O^*(n^{4/3})$
time for $d=3,4$, and $O^*\left(n^{\frac{2\lceil d/2 \rceil}{1+\lceil d/2 \rceil}}\right)$
time for $d\ge 5$. For $d\ge 5$, the maximum matching in the sparsified graph is 
computed by the algorithm of Micali and Vazirini \cite{MV} in $O(n^{3/2})$ time,
so the sparsification cost dominates the cost of the algorithm.)

For example, in $d=4$ dimensions, we can compute the smallest $r^*$ such that $G(r^*)$ 
contains a perfect matching (equivalently, a bottleneck matching $M$ of the set of the 
centers of the disks of $\U$) in $O^*(n^{17/11})$ expected time. 
(Here $t=4$ and $D(n)=O(n^{3/2})$.)
\matya{This is not interesting if there exists, as in the planar case, a linear-size graph 
that contains $M$ and that can be computed fast.}
For $d\ge5$, $t=d$ and $D(n)$ is determined by the running time of the sparsification, 
which is $O^*(n^{3/2})$ for $d=5,6$. 

\matya{I think that their results for fat axis-parallel boxes of similar size in $\reals^d$ 
are irrelevant, since selection in this case is efficient. In particular, we can compute 
the $k$'th critical value for growing segments on the line in near-linear time, right?}

\matya{The same goes for translates of a convex object in the plane --- irrelevant.}

\micha{I think you are right on both counts.}
}


\medskip
\noindent{\bf (vi) Generalized distance selection problems.}
The next kind of applications that we consider are generalizations of the distance selection 
problem in the plane. Let $\S$ be a set of $n$ pairwise disjoint segments in the plane. 
We define the distance between a pair of segments as the minimum value $r$ such that, 
if we expand the length of each of the segments about its center by $r$, then they intersect. 
The expansion can be either additive (i.e., we add $r$ to each of the segment lengths) or 
multiplicative (i.e., we multiply each of the segment lengths by $r$). We want to find the 
$k$'th smallest distance among the $n \choose 2$ pairwise distances determined by the segments 
in $\S$, for a given parameter $k$. This is equivalent to finding the smallest $r$ for which 
there are exactly $k$ vertices (i.e., intersection points) in the arrangement of $\S_r$, where 
$S_r$ denotes the set of the segments of $\S$, each expanded by $r$.

On the one hand, this problem can be solved, using more standard techniques 
(that we discuss later in the paper), in 
$O^*(n^{8/5})$ time. On the other hand, the corresponding decision problem --- given $r$, 
is the number of vertices in the arrangement of $\S_r$ greater than, equal to, or smaller 
than $k$ ---  can be solved, using line sweeping, in $O((n+k)\log n)$ time. Our technique 
allows us to `combine' the two results and obtain a solution to the selection problem, 
which is more efficient than the best known one, provided that $k$ is not too large. 
More precisely, we can find the $k$'th smallest distance in $\S$ in $O^*(n^{8/11}k^{6/11})$
randomized expected time, which is more efficient than the standard solution when $k = o(n^{8/5})$. 

As another example, consider a set $\D$ of $n$ pairwise disjoint disks of arbitrary radii 
in the plane. We seek the smallest $r$, such that, if we expand each of the disks in $\D$ 
by $r$ (i.e., we either add $r$ to each of the radii, or multiply each of the radii by $r$), 
we get $k$ intersecting pairs of disks, for some pre-specified integer $1 \le k \le {n \choose 2}$. 
In this case, the problem can be solved using more standard techniques (that we discuss later) 
in $O^*(n^{3/2})$ time, and the 
corresponding decision problem can be solved using (a careful implementation) of the line-sweeping 
technique in $O((n+k)\log n)$ time. Again, our technique allows us to `combine' these results and 
obtain an algorithm that finds the smallest $r$ such that there are $k$ intersecting pairs in 
$\D_r$, where $\D_r$ is the set of the disks in $\D$, each expanded by $r$.
The algorithm runs in $O^*(n^{3/4}k^{1/2})$ randomized expected time, which is faster
than the standard solution when $k = o(n^{3/2})$.

\medskip
\noindent{\bf (vii) The maximum-height independent towers problem.} 
Finally, in Section~\ref{sec:mpl} we consider the \emph{maximum-height independent towers problem}, 
in which we are given an $x$-monotone polygonal line $T$ with $n$ vertices, which we think of as a
\emph{1.5-dimensional terrain}, and a set $Q$ of $m$ points on $T$, and the goal is to find
the maximum height $h^*$ such that if we erect a vertical tower of height $h^*$ at each point
of $Q$, the tips of the towers are mutually invisible (i.e., no segment connecting any two of them 
lies fully above $T$). This problem has a decision procedure (given $h$, determine whether 
all the tower tips are mutually invisible), due to Ben Moshe et al.~\cite{BHKM}, that runs
in near-linear time. Using this procedure, we show that the optimization problem can 
be solved in $O^*(n^{6/5})$ randomized expected time.

The application of our technique to the maximum-height independent towers problem is different
in that the critical values of the parameter that we want to optimize are determined
by \emph{triples} of input points (each of these values is the vertical distance from a point 
to the segment connecting two other points). This calls for certain modifications
of the interval shrinking subprocedure, which we also present in Section~\ref{sec:mpl}.



\section{The underlying machinery: Interval shrinking and \\ bifurcation trees} \label{sec:prelim}

The material in this section provides the infrastructure, in rather full generality, on which
our algorithms are based. 

\subsection{Shrinking the interval of critical values} \label{sec:shrink}

\medskip
\noindent{\bf Distances in a planar point set.}
We first recall the earlier result of \cite{BFKKS}, which handles the case of 
distances for planar point sets.\footnote{%
  The bounds stated below are slightly different from those in \cite{BFKKS},
  which only considered `bichromatic' distances between points in two input sets.}
We have a set $P$ of $n$ points in the plane, and a parameter $L \ll \binom{n}{2}$.
The goal is to find an interval $(\alpha,\beta]\subset\reals$ that 
(a) contains the optimum value $r^*$, and (b) contains at most $L$ distances determined by $P$.

\begin{theorem}[Ben Avraham et al.~\protect{\cite{BFKKS}}] \label{th:shrink2d}
Let $P$ and $L$ be as above. We can construct an interval 
$I = (\alpha,\beta]\subset \reals$, such that $I$ contains the optimum value $r^*$,
and $I$ contains at most $L$ distances determined by the points of $P$. The expected running
time is $O^*(n^{4/3}/L^{1/3} + D(n))$, where $D(n)$ is the cost of the decision procedure.
\end{theorem}

\medskip
\noindent{\bf Remark.}
The term $O^*(D(n))$ in the above bound replaces the term $O(n\log n)$ in \cite{BFKKS}. 
It comes from applying the decision procedure a logarithmic number of times in the interval
shrinking algorithm. The cost of these applications is $O(n\log n)$ in \cite{BFKKS} but 
in general it is $O(D(n)\log n) = O^*(D(n))$. 

\medskip
\noindent{\bf The general case of interval shrinking.}
%
In the general case we have an input set $S$ of $n$ semi-algebraic objects of constant complexity
in any dimension $\reals^d$ (e.g., in the plane we may have segments, disks, ellipses, etc.),
and a semi-algebraic predicate $\Pi(o,o';r)$ of constant complexity, over pairs $o,o'$ of 
objects of $S$, which is monotone in an additional `growth parameter' $r$. Each pair $o,o'$ 
defines a \emph{critical value} $r_{o,o'}$, which is the minimum $r$ for which $\Pi(o,o';r)$ 
is true. (As already mentioned, the minimum indeed exists in all our applications.)

The general \emph{interval-shrinking} problem goes as follows. We are given a parameter 
$L \ll \binom{n}{2}$ and seek an interval $I = (\alpha,\beta]\subset \reals$, such that 
$I$ contains the optimum value $r^*$, and $I$ contains at most $L$ critical values
determined by the objects of $S$.

The interval-shrinking procedure of \cite{BFKKS} is related to the classical 
\emph{distance selection} algorithm of \cite{AASS,KS}, and our general procedure is
similarly related to a corresponding \emph{critical-value selection} problem, in which we want 
to compute the $k$-th smallest critical value determined by pairs of objects of $S$. 
This problem is of independent interest, and can be solved using a suitable
simpler variant of the technique presented below, as will be noted later. 

The general interval-shrinking problem can be solved by adapting and extending the
high-level approach in \cite{BFKKS}. The technical details, though, are different and
more involved, as they rely on recent techniques for semi-algebraic range searching. 
Specifically, let $t$ be the number of degrees of freedom of the objects in $S$,
namely the number of real parameters needed to specify an object. In the generalization
of the main step of the procedure of \cite{BFKKS}, we have an interval $I = (r_1,r_2]$,
known to contain $r^*$, and we want to determine whether it contains at most $L$ 
critical values. If this is the case we terminate the procedure and return $I$.
If not, we want to shrink $I$ into a subinterval $I'$ that contains at most some
fixed fraction of the number of the critical values in $I$ (while still containing $r^*$).

To perform this step we reduce the problem to the following problem of batched 
range searching with semi-algebraic ranges. We represent the objects of $S$ as 
$n$ points in $\reals^t$, denoting the resulting set of points as $P$,
and also map each object $o'\in S$ to the range
\[
\sigma_{o'} = \{ u\in \reals^t \mid r_1 < r_{u,o'} \le r_2 \} ,
\]
which, by our assumptions, is semi-algebraic of constant complexity.\footnote{%
  Here $u$ ia an arbitrary point in $\reals^t$, designating an arbitrary possible range, 
  not necessarily in $S$. This reduction still works even when we let $\sigma_{o'}$
  include points $u$ for which there is no actual range that $u$ designates.}
Let $\Sigma$ denote the resulting collection of ranges. We have $|P| = |\Sigma| = n$, 
but it will be more convenient to consider a bipartite version of the problem, in 
which $P$ and $\Sigma$ come from different subsets of $S$ and may have different sizes. 
Put $m = |P|$ and $n = |\Sigma|$. Note also that the problem has a symmetric dual 
version, in which we map the ranges of $\Sigma$ to points in $\reals^t$, and the 
points of $P$ to constant-complexity semi-algebraic ranges in $\reals^t$. 
Both versions will be used in the algorithm.

\medskip
\noindent{\bf The semi-algebraic range-searching mechanism.} 
We adapt the range searching machinery of \cite{MP}
(or of \cite{AAEZ}), but, in order to gain efficiency, we need to terminate the 
hierarchical process prematurely. For this we have to go into the inner workings
of the technique of \cite{MP}. Concretely, we recall the following main technical 
result of \cite{MP} (where the notations have been changed to avoid overloading 
of notations introduced earlier and to conform to our setup):

\begin{theorem}[Matou\v{s}ek and Pat\'akov\'a~\protect{\cite[Theorem 1.1]{MP}}] \label{thm:mp}
For every integer $t > 1$ there is a constant $K$ such that the following hold.
Given an $n$-point set $P \subset \reals^t$ and a parameter $s > 1$, there are numbers 
$s_1,s_2,\ldots,s_t \in [s, s^K]$, positive integers $\xi_1,\xi_2,\ldots,\xi_t$, a partition
\[
P = P^* \cup \bigcup_{i=1}^t \bigcup_{j=1}^{\xi_i} P_{ij}
\]
of $P$ into disjoint subsets, and for every $i,j$, a connected set $S_{ij} \subseteq \reals^t$ 
containing $P_{ij}$, such that $|P_{ij}| \le n/s_i$ for all $i, j$, $|P^*| \le s^K$, 
and the following holds:

If $h \in \reals[x_1, \ldots, x_t]$ is a polynomial of degree bounded by a constant $D_0$, 
and $X = Z(h)$ is its zero set, then, for every $i = 1,2,\ldots,t$, the number of the sets
$S_{ij}$ crossed by (intersected by but not contained in) $X$ is at most 
$O\left(s_i^{1-1/t}\right)$, with the implicit constant also depending on $D_0$.
\end{theorem}

The interval-shrinking procedure is carried out as follows. We fix a constant parameter $s$.
If $m \ge n$ we apply Theorem~\ref{thm:mp} to $P\subset \reals^t$, and if $m \le n$ we apply 
Theorem~\ref{thm:mp} to the set $\Sigma^*$ of points dual to the ranges of $\Sigma$. 
Assume without loss of generality that $m\ge n$, and follow the notations in the theorem.

For each $i = 1,2,\ldots,t$, each of the boundary surfaces of the ranges in $\Sigma$ 
crosses at most $O\left(s_i^{1-1/t}\right)$ sets $S_{ij}$. Denote by $\Sigma_{ij}$ 
(resp., $\Sigma^0_{ij}$) the subset of ranges whose boundary surfaces cross $S_{ij}$ 
(resp., that fully contain $S_{ij}$, and thus $P_{ij}$). Put $q_{ij} = |\Sigma_{ij}|$.
Then we have, for each $i$,
$\sum_{j=1}^{\xi_i} q_{ij} = O\left(n s_i^{1-1/t}\right)$.
For each $i=1,\ldots,t$ and each $j=1,\ldots,\xi_i$, we (implicitly) form the biclique 
$P_{ij}\times \Sigma^0_{ij}$, and output its vertex sets, and face a subproblem involving 
$P_{ij}$ and $\Sigma_{ij}$, of respective sizes at most $n/s_i$ and $q_{ij}$, which we solve
recursively, possibly switching to the dual setup, depending on which of these two sizes 
is larger. We stop the recursion when both of the sizes become smaller than $L$. At the 
bottom of recursion, we also output (the vertex sets of) the corresponding bicliques 
$P_{ij}\times \Sigma_{ij}$. 
We note that the first kind of bicliques $P_{ij}\times \Sigma^0_{ij}$ has the property 
that every edge $(o,o')$ in any such biclique satisfies $r_1 < r_{o,o'} \le r_2$, but
this is not necessarily the case for the second kind of bicliques $P_{ij}\times \Sigma_{ij}$.
This is the setup that the machinery in \cite{BFKKS} processes.

The efficiency of the procedure is determined by two comparable quantities: the overall
size $K(m,n)$ of the vertex sets of the bicliques that we output, and the time $T(m,n)$
used by the procedure. It follows from the definition of the procedure and from 
Theorem~\ref{thm:mp} that the two quantities are asymptotically the same, so we will
analyze just one of them, say $K(m,n)$.

\begin{lemma} \label{lem:mprec}
For any $\eps>0$ there is a constant $A$ that depends on $\eps$, such that, for any set
of at most $m$ points in $\reals^t$ and any set of at most $n$ ranges in $\reals^t$,
of the type considered above, we have
\begin{equation} \label{eq:mprec}
K(m,n) \le A \left( \frac{m^{t/(t+1)+\eps} n^{t/(t+1)+\eps}}{L^{(t-1)/(t+1)}} + m^{1+\eps} + n^{1+\eps} \right) .
\end{equation}
\end{lemma}

\noindent
{\bf Proof.}
The proof is by induction. The base case $m,n \le L$ is trivial since then $K(m,n) = O(m+n)$
(this is the overall size of the vertex sets of the second kind of bicliques that the procedure
outputs; the linear bound follows since $s$ is assumed to be constant---see below).
Consider then the case where, say, $m \ge n$ and $m > L$, and assume that (\ref{eq:mprec})
holds for all smaller values $m'\le m$, $n' < n$ and $m' < m$, $n' \le n$.
Apply Theorem~\ref{thm:mp} to the primal setup (we would apply it in the dual in the 
complementary case where $n\ge m$ and $n > L$), and use the notations in the theorem.
Since $s$ is a constant, the nonrecursive cost of the procedure is at most $B(m+n)$, 
where $B$ is a constant that depends on $s$ and the various other constant parameters 
of the setup. This also takes care of processing $P^*$, which is of constant size.
By induction hypothesis, for each $i$ and $j$, the cost of the recursive processing
of $P_{ij}$ and $\Sigma_{ij}$ is at most
\begin{equation} \label{eq:ind}
A\left( \frac{p_{ij}^{t/(t+1)+\eps} q_{ij}^{t/(t+1)+\eps}}{L^{(t-1)/(t+1)}} + p_{ij}^{1+\eps} + q_{ij}^{1+\eps} \right) ,
\end{equation}
where $p_{ij} = |P_{ij}|$. Observe that we have $p_{ij} \le m/s_i$ for each $j$, and the 
quantities $m_i := \sum_{j=1}^{\xi_i} p_{ij}$ satisfy $\sum_{i=1}^d m_i \le m$ (since the
decomposition in Theorem~\ref{thm:mp} is into disjoint subsets). Recall also that 
$\sum_{j=1}^{\xi_i} q_{ij} \le cn s_i^{1-1/t}$ for every $i$ and for a suitable absolute constant $c$.

We now sum the bounds in (\ref{eq:ind}) over $j$ for each fixed $i$. 
Using H\"older's inequality, the sum is upper bounded by
\begin{gather*}
A\left( \sum_{j=1}^{\xi_i} \frac{p_{ij}^{t/(t+1)+\eps} q_{ij}^{t/(t+1)+\eps}}{L^{(t-1)/(t+1)}} + 
\sum_{j=1}^{\xi_i} p_{ij}^{1+\eps} + \sum_{j=1}^{\xi_i} q_{ij}^{1+\eps} \right) \\
\le A\left( (m/s_i)^{(t-1)/(t+1)+2\eps} 
\sum_{j=1}^{\xi_i} \frac{p_{ij}^{1/(t+1)-\eps} q_{ij}^{t/(t+1)+\eps}}{L^{(t-1)/(t+1)}} + m_i^{1+\eps} + 
(cn s_i^{1-1/t})^{1+\eps} \right) \\
\le A\left( \frac{(m/s_i)^{(t-1)/(t+1)+2\eps}}{L^{(t-1)/(t+1)}} m^{1/(t+1)-\eps} 
\left( cn s_i^{1-1/t} \right)^{t/(t+1)+\eps} + m_i^{1+\eps} +
(cn s_i^{1-1/t})^{1+\eps} \right) \\
= A\left( \frac{m^{t/(t+1)+\eps}n^{t/(t+1)+\eps}}{L^{(t-1)/(t+1)}} \cdot 
\frac{c^{t/(t+1)+\eps} }{s_i^{(1+1/t)\eps} }
+ m_i^{1+\eps} + (cs_i^{1-1/t})^{1+\eps}n^{1+\eps} \right) .
\end{gather*}
Assuming that $s$ (and thus each $s_i$) is chosen sufficiently large, we may assume that
the factor $\frac{c^{t/(t+1)+\eps} }{s_i^{(1+1/t)\eps} }$ is smaller than $1/(4d)$.
The problematic factor is $Q := (cs_i^{1-1/t})^{1+\eps}$ in the last term, 
which in general will be larger than $1$, making the induction step problematic. To address 
this issue, one can easily verify that, under the assumption that $m \ge n$ and $m > L$, we have
\[
n \le \frac{m^{t/(t+1)}n^{t/(t+1)}}{L^{(t-1)/(t+1)}} ,\qquad\text{or}\qquad 
n^{1+\eps} \le \frac{m^{t(1+\eps)/(t+1)}n^{t(1+\eps)/(t+1)}}{L^{(t-1)(1+\eps)/(t+1)}} .
\]
This allows us to write
\begin{equation} \label{eq:replin}
AQn^{1+\eps} \le 
A \frac{m^{t/(t+1)+\eps}n^{t/(t+1)+\eps}}{L^{(t-1)/(t+1)}} \cdot 
\frac{Q}{ m^{\eps/(t+1)} n^{\eps/(t+1)} L^{(t-1)\eps/(t+1)}} ,
\end{equation}
and we can make the second factor smaller than $1/(4d)$ by assuming that $m$, $n$ and $L$ are sufficiently large.
(When $m$ and $n$ are small, (\ref{eq:mprec}) will hold by choosing $A$ sufficiently large.)

We now sum up the modified bounds (in which the inequality in (\ref{eq:replin}) is substituted
into the bound) over $i=1,\ldots,d$, and obtain the overall bound
\[
A \sum_{i=1}^d
\left( \frac{m^{t/(t+1)+\eps}n^{t/(t+1)+\eps}}{L^{(t-1)/(t+1)}} \cdot \left( \frac{1}{4d} + \frac{1}{4d} \right)
+ m_i^{1+\eps} \right) 
\le \frac{A}{2} \cdot \frac{m^{t/(t+1)+\eps}n^{t/(t+1)+\eps}}{L^{(t-1)/(t+1)}} + Am^{1+\eps} .
\]
This, and the similar analysis of the symmetric case $n\ge m$ and $n > L$, 
establish the induction step and thus complete the proof of the lemma.
$\Box$

The lemma implies the following main result of this section.
\begin{theorem} \label{th:shrink}
Given a set $S$ of $n$ semi-algebraic geometric objects of constant complexity (in any dimension)
with $t$ degrees of freedom,
a growth parameter $r$ and an $r$-monotone predicate $\Pi(o,o';r)$ over pairs $o,o'\in S$,
and a parameter $n\le L\ll \binom{n}{2}$, we can find an interval $I$ that contains the optimum 
value $r^*$ and contains at most $L$ critical values determined by pairs of objects in $S$, 
in randomized expected time $O^*(n^{2t/(t+1)}/L^{(t-1)/(t+1)} + D(n))$.
\end{theorem}

\medskip
\noindent{\bf Remark.}
A simplified version of the above machinery, in which we run the range searching algorithm 
to completion, yields an algorithm for \emph{critical value selection}, in which we want
to compute the $k$th smallest critical value among those determined by a set of $n$ objects, 
as defined above, that runs in randomized expected time $O^*(n^{2t/(t+1)})$. 

\medskip
\noindent{\bf Shrinking the interval of critical distances in three and higher dimensions.} 
As an illustration of the general machinery described above, consider the case where the
objects are points in $\reals^d$, for any $d\ge 3$, and the critical values are the
distances between the points.
In this special case the problem involves range searching in $t=d$ dimensions; 
the ranges are actually spherical shells with inner radius $r_1$ and outer radius $r_2$. 
Theorem~\ref{th:shrink} then implies:

\begin{theorem} \label{th:shrinkdd}
Given a set $P$ of $n$ points in $\reals^d$, for any $d\ge 2$,
and a parameter $n\le L\ll \binom{n}{2}$, we can find an interval 
$I$ that contains the optimum value $r^*$ and contains at most $L$ 
distances determined by $P$, in randomized expected time $O^*(n^{2d/(d+1)}/L^{(d-1)/(d+1)} + D(n))$. 

\noindent
In particular, for $d=3$ the expected running time is $O^*(n^{3/2}/L^{1/2} + D(n))$.
\end{theorem}

\medskip
\noindent{\bf Remark.}
We have focused here on examples where the critical values are inter-point distances,
but the machinery in Theorem~\ref{th:shrink} is clearly much more widely applicable, and
several of our applications, given in subsequent sections, will use the theorem in more
general contexts.


\subsection{Bifurcation-tree construction} \label{sec:bifur}

We now consider, in fairly full generality, the second step of the algorithm, which
simulates the decision procedure using bifurcation trees. This part is similar to the
corresponding procedure in \cite{BFKKS}, but requires several modifications to make it 
more broadly applicable. 

Consider an optimization problem {\sc Opt} that involves a set $S$ of $n$ objects
as above, and seeks the minimum value $r^*$ of some 
real parameter $r$ at which some property (that is monotone in $r$) holds for 
the set of critical values determined by pairs of objects of $S$, with respect
to an $r$-monotone predicate $\Pi(o,o';r)$ as defined above. We assume that the decision 
procedure accesses its parameter $r$ only through comparison tests against concrete values. 
In particular, this is how the simulation of the decision procedure at the unknown $r^*$
accesses $r^*$. Let $D(n)$ denote an upper bound on 
the number of such comparisons in any execution of the decision procedure.
We also assume that the overall running time of the decision procedure is $O(D(n))$.

Assume that we have already applied the interval-shrinking procedure
for some value $L$ that will be determined shortly,
and denote by $I$ the resulting shrunken interval. We choose another parameter $s$, 
and simulate the execution of the decision procedure at (the unknown) $r^*$. 
At each comparison test that we encounter, with some concrete value $r_0$, we check 
whether $r_0$ lies inside or outside $I$. If $r_0$ lies outside $I$ we know the result 
of the comparison (because $r^*\in I$), and follow the unique way in which the simulation of 
the decision procedure at $r^*$ proceeds; we denote this step as an \emph{outdegree-one} step. 
If $r_0$ lies inside $I$, we bifurcate, following an execution path at which $r^* < r_0$, 
and a path at which $r^* > r_0$, referring to this step as an \emph{outdegree-two} step.
There is also the possibility that $r^* = r_0$, but in this case (if it is the correct one)
the decision procedure terminates, so this case does not cause any further expansion of 
the tree beyond this test.

We continue to expand the resulting \emph{bifurcation tree} $T$ in this manner. 
We stop each branch of the tree (a path of the tree that starts at a bifurcation node
and consists of only outdegree-one nodes)
when it accumulates $s$ nodes beyond its bifurcation node. If we reach another
bifurcation before encountering $s$ outdegree-one nodes, we bifurcate there and start
the count afresh for each of the two outgoing paths. We stop the entire construction 
of the tree either when each of its terminal branches contains $s$ outdegree-one nodes,
or when the tree comes to contain $D(n)$ nodes, whichever happens earlier. 
When we stop the construction, we end a phase of the algorithm.
At that time we collect the outdegree-two nodes of $T$ that we have created, and run a
binary search through the sequence of the corresponding critical values of $r$, using 
the unsimulated decision procedure to guide the search. This yields two consecutive
critical $r$-values $r_1$, $r_2\in I$ that enclose $r^*$. This allows us to continue the 
simulation of the decision procedure, in a unique manner, from its state at the root 
of $T$ to its state at the suitable leaf. We also replace $I$ by the smaller interval
$(r_1,r_2)$, which still contains $r^*$. We then start a new phase, executed in the 
same manner, from the final state of the previous phase, and with the new, further
shrunken interval $I$, and continue to do so until we complete the simulation of the 
decision procedure, or, more precisely, when we reach a comparison that has
$r^*$ as a critical value, an event that must occur during the simulation.

Suppose that we stop because of the first condition, namely each branch of the tree 
that ends at a leaf contains $s$ outdegree-one nodes; this is a `successful' phase. 
In particular, this implies that, for every value in $I$ encountered in this phase, 
the decision procedure has made at least $s$ comparisons during this phase with 
critical values outside $I$. It follows that the simulated procedure will run to 
completion after at most $D(n)/s$ successful phases. Since the cost of each phase 
is $O(D(n)\log n)$ (to construct the tree and to run the binary search through 
its bifurcations), the overall cost of the successful phases is
${\displaystyle O\left( \frac{D^2(n)\log n}{s}\right)}$.

Consider next the case where we stop because of the second condition, namely $T$ 
contains $D(n)$ nodes; this is an `unsuccessful' phase. If we denote by $x$ the number 
of outdegree-two nodes in $T$, then the length of each branch of the tree between 
two consecutive outdegree-two nodes is at most $s$, so $D(n)\le xs$, or $x\ge D(n)/s$.
In other words, in an unsuccessful phase we encounter at least $D(n)/s$ critical values 
in $I$, and the binary search through them will shrink $I$ further, to a subinterval that
contains none of these values (except for its two endpoints). Hence the number
of unsuccessful phases is at most $L/(D(n)/s) = Ls/D(n)$, and their overall cost is
${\displaystyle O\left(\frac{LsD(n)\log n}{D(n)}\right) = O\left(Ls\log n\right)}$.

We balance these two bounds by choosing $s = D(n)/L^{1/2}$, noting that this value
of $s$ is at least 1, and thereby obtain the following proposition.

\begin{proposition} \label{pr:bifur}
The overall cost of the bifurcation tree portion of the algorithm, following
a suitable interval-shrinking step that produces an interval containing $r^*$
and at most $L$ critical values of $r$, is $O\left(L^{1/2}D(n)\log n\right)$.
\end{proposition} 

Note that this part of the algorithm is fairly general and is independent of the
specific problem at hand. All it requires is that we have an interval $(\alpha,\beta]$
that is known to contain at most some number $L$ of critical values. (It also assumes,
or rather requires, that the decision procedure accesses the parameter $r$ only via 
comparisons, which can also be sign tests of constant-degree polynomials in $r$.)

\subsection{The overall algorithm} \label{sec:overall}

We now balance the cost of the bifurcation-tree part with that of the interval-shrinking part.
The actual balancing depends on the number $t$ of degrees of freedom of the input objects. 
For problems involving distances between points in the plane, the overall
cost is $O^*(n^{4/3}/L^{1/3} + L^{1/2}D(n))$,
so we choose $L^{5/6} = n^{4/3}/D(n)$, or $L = n^{8/5}/D(n)^{6/5}$, making the overall 
cost of the algorithm $O^*(n^{4/5}D(n)^{2/5})$.

When the decision procedure takes (near) linear time, the running 
time is $O^*(n^{6/5})$. The new bound lies strictly between $D(n)$
and $O^*(n^{4/3})$ for $D(n) = o(n^{4/3})$ and $D(n) = \omega(n)$.

For problems involving distances in three dimensions, the cost of the interval-shrinking procedure is
$O^*(n^{3/2}/L^{1/2})$ (see Theorem~\ref{th:shrinkdd}), so the balancing leads to choosing 
$L = n^{3/2}/D(n)$, making the overall cost of the algorithm $O^*(n^{3/4}D(n)^{1/2})$. 
For linear-time decision procedures, this becomes $O^*(n^{5/4})$.
In full generality, we have:
\begin{theorem} \label{th:main}
Let {\sc Opt} be an optimization problem on a set $S$ of $n$ semi-algebraic objects
of constant complexity, in any dimension, which have $t$ degrees of freedom. Assume
that {\sc Opt} depends on critical values determined by pairs of objects of $S$ and 
a growth parameter $r$, in terms of an $r$-monotone predicate $\Pi(o,o';r)$ as defined above.
Assume further that {\sc Opt} has a decision procedure that is based (solely) on comparisons 
involving critical values determined by pairs of objects of $S$, so that it uses 
at most $D(n)$ such comparisons, and runs in overall $O(D(n))$ time. 

\medskip
\noindent
We can then solve {\sc Opt} by an algorithm that runs in
$O^*\left(n^{2t/(3t-1)}D(n)^{(2t-2)/(3t-1)}\right)$ randomized expected time.
\old{
\micha{Perhaps remove the rest of this para?}
In particular, this becomes $O^*(n^{4/5}D(n)^{2/5})$ randomized expected time 
when the critical values are distances between points in the plane, 
$O^*(n^{3/4}D(n)^{1/2})$ randomized expected time for distances between points in three dimensions, and, in
arbitrary dimension $d$, $O^*\left(n^{2d/(3d-1)}D(n)^{(2d-2)/(3d-1)}\right)$ randomized expected time.
For decision procedures with (near) linear running time, the bound becomes
$O^*(n^{(4t-2)/(3t-1)})$, which, for distances between points becomes
$O^*(n^{6/5})$, $O^*(n^{5/4})$, and $O^*(n^{(4d-2)/(3d-1)})$, respectively.
}
\end{theorem}


\section{Applications}

\subsection{The reverse shortest path problem for unit-disk graphs} \label{sec:sp2d}

\begin{figure}[ht]
\centering
\includegraphics[scale=0.8]{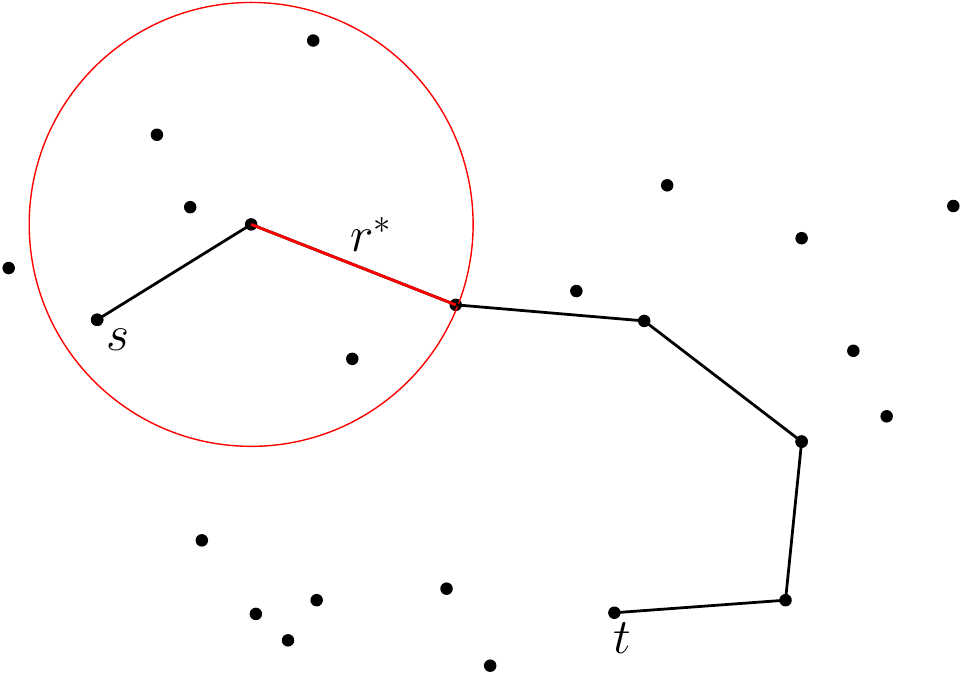}
\caption{\sf The smallest value $r^*$ for which $G(r^*)$ has a path between $s$ and $t$ of length at most 6.}
\label{fig:reverse}
\end{figure}

In this problem we are given a set $P$ of $n$ points in the plane, two points
$s,t\in P$, and a parameter $k\le n-1$, and the goal is to compute the smallest
value $r^*$ so that there exists a path between $s$ and $t$ of length at most
$k$ in the graph $G(r^*)$, which is the graph over $P$ whose edges are all the 
pairs $p,q\in P$ such that $\dist(p,q) \le r^*$; see Figure~\ref{fig:reverse}.


This problem has recently been studied in Wang and Zhao~\cite{WZ}, who presented a solution
that runs in $O^*(n^{5/4})$ time. Wang and Zhao use a decision procedure for this problem, 
due to Chan and Skrepetos~\cite{CS}, where $r$ is specified and the goal is to determine 
whether $G(r)$ contains a path from $s$ to $t$ of length at most $k$. The procedure is 
based on a careful execution of BFS in $G(r)$, and runs in $D(n) = O(n)$ time (after 
an initial sorting step). 

Chan and Skrepetos' procedure constructs a uniform grid of cell size $r/\sqrt{2}$, where 
$r$ is the input parameter, and distributes the points of $P$ among the grid cells.
Na\"ively implemented (using the floor function, say), this step does not conform to 
our requirements, that the data be accessed only via comparisons. Nevertheless we can
distribute the points of $P$ using simple binary search. To do so, let $a$ denote the 
distance $\dist(s,t)$. Note that $r^*\le a$, because 
$G(a)$ has a single edge connecting $s$ to $t$. We also have $r^*\ge a/k$,
because for smaller values of $r^*$ there is no path of length $k$ in $G(r^*)$
that connects $s$ and $t$ (this is a consequence of the triangle inequality). 
We therefore have $a/k\le r^*\le a$. Assume then that $r$ satisfies these 
inequalities (other values of $r$ can be rejected right away), and observe
that every point on a $k$-path from $s$ to $t$ in $G(r)$ must be at distance
at most $\Theta(ak)$ from $s$. This implies that any such path must be fully
contained in a subgrid of size $O(k^2)\times O(k^2)$ around $s$. Hence, using
binary search through the $x$- and $y$-coordinates of this subgrid we can
distribute the relevant points of $P$ in the cells of that subgrid, using only
comparisons.

Plugging this into our mechanism, we obtain
\begin{theorem}
The reverse shortest path problem for unit-disk graphs, as just formulated,
can be solved in $O^*(n^{6/5})$ randomized expected time.
\end{theorem}

By setting $k=n-1$ we get the following corollary, which is of independent interest.
\begin{corollary}
Given $P$ and $s,t \in P$, as above, one can find the smallest value $r^*$ so that 
there exists a path between $s$ and $t$ in $G(r^*)$, in $O^*(n^{6/5})$ randomized expected time.
\end{corollary}

\subsection{The reverse weighted shortest path problem for unit-disk
graphs} \label{sec:wsp2d}

In the weighted variant of the reverse shortest path problem, we
want to find the minimum value $r^*$ such that $G(r^*)$, defined as above,
contains a path from $s$ to $t$ of length at most $w$, where each edge
of $G(r^*)$ has a weight, equal to its Euclidean distance, and the
length of the path is the sum of the weights of its edges. 

The decision procedure for this problem has been studied by Wang and Xue~\cite{WangX20},
who showed that, for a given $r$ and a source point $s$, all shortest paths in $G(r)$
from $s$ to the other points of $P$ can be computed in $O(n\log^2n)$ time.

As in the unweighted case, the decision procedure constructs a uniform 
grid of cell size $r/2$, and distributes the points of $P$ 
among the grid cells. We modify the procedure as we did in the 
unweighted case, and then apply our machinery to the modified procedure.
More precisely, we now have $a\le w$ (by the triangle inequality), and 
$\frac{a}{n} \le r^* \le a\le w$, where the left inequality is again a
consequence of the triangle inequality. For any $\frac{a}{n} \le r \le a$, 
any point on a path from $s$ to $t$ in $G(r)$ must lie at distance at most
$an$ from $s$, which implies that any such path is fully contained in an 
$O(n^2)\times O(n^2)$ grid around $s$, and the same argument as before,
replacing $k$ by $n$, lets us distribute the points of $P$ in this 
subgrid by binary search, using only comparisons.

Plugging this into our mechanism, we obtain
\begin{theorem}
The reverse weighted shortest path problem for unit-disk graphs, as just 
formulated, can be solved in $O^*(n^{6/5})$ randomized expected time.
\end{theorem}

\old{
\subsection{Minimum bottleneck moving spanning tree} \label{sec:moving_sp}

\micha{I did not have time to digest this in depth. We can talk about it at some point,
but for SODA I suggest that we put it in the boydem. It is indeed commented out for now.}


Consider a set $P$ of $n$ moving points in the plane, where each point $p \in P$ is moving along a line $l_p$, in constant velocity $v_p$, from an initial location $p(0)$ (its location at time $t=0$) to a final location $p(1)$ (its location at time $t=1$). In general, we denote the location of $p$ at time $t \in [0,1]$ by $p(t)$. Let $T$ be a tree over $P$, i.e., each of $T$'s nodes represents a unique point of $P$. For $t \in [0,1]$, we denote by $T(t)$ the spanning tree over $P(t)=\{p_1(t),\ldots,p_n(t)\}$ that is obtained from $T$; that is, the edges $(p_i,p_j)$ of $T$ are drawn as line segments $p_i(t)p_j(t)$ in $T(t)$. The \emph{bottleneck} $b(T(t))$ of $T(t)$ is the Euclidean length of a longest edge in $T(t)$, and the \emph{bottleneck} of $T$ is $b(T) = \max_{t \in [0,1]} b(T(t))$. In the minimum bottleneck moving spanning tree problem, the goal is to compute a tree $T^*$ over $P$ with minimum bottleneck. 

This problem was introduced by Akitaya et al.~\cite{AkitayaBBCMSS21} (motivated by considerations concerning the visualization of time-varying spatial data), who presented an $O(n^2)$-time algorithm for computing $T^*$. This result was recently improved by Wang and Zhao~\cite{WZ2}, who gave an $O(n^{4/3}\log^3 n)$-time algorithm for computing $T^*$. We show below that it can be further improved to $O^*(n^{6/5})$ by applying our machinery.

The previous solutions, as well as ours, rely on the key observation (made in~\cite{AkitayaBBCMSS21}) that for any tree $T$ over $P$, the bottleneck of $T$ is either $b(T(0))$ or $b(T(1))$, or more precisely, $b(T) = \max\{b(T(0)),b(T(1))\}$, which implies that $b(T) \in \{(\dist(p_i(0),p_j(0)) \mid p_i,p_j \in P\} \cup \{\dist(p_i(1),p_j(1)) \mid p_i,p_j \in P\}$. The algorithm of Wang and Zhao performs a binary search in this set of candidate values, using an $O^*(n^{4/3})$-time distance selection procedure to produce the next candidate value $r$ to be compared with $b(T^*)$. The comparison itself is resolved as follows. Compute a biclique representation of the unit disk graph $G_0(r)$, which is the graph over $P(0)$ whose edges are all the pairs $p_i(0),p_j(0) \in P(0)$ such that $\dist(p_i(0),p_j(0)) \le r$. Next, perform a BFS in $G_0(r)$ from an arbitrary point $p(0)$, using the biclique representation. The BFS attempts to construct the shortest path tree rooted at $p(0)$ by levels, beginning from level~0 which is $\{p(0)\}$. However, when a new level is constructed from the previous one, retain only the points $p_j(0)$ of the new level for which there exists a point $p_i(0)$ of the previous level such that $\dist(p_i(1),p_j(1)) \le r$. Finally, $b(T^*) \le r$, if and only if, the resulting tree is a spanning tree of $P(0)$.

Wang and Zhao show that the decision procedure described above can be implemented in $O^*(n^{4/3})$ time. For this, they prove the following useful theorem:
\begin{theorem}[Wang and Zhao~\protect{\cite{WZ2}}]
\label{thm:WZ}
Given a value $r$ and a set $Q$ of $n$ points in the plane, one can build a linear-size data structure in $O(n \log n)$ time such that the following first operation can be performed in $O(\log n)$ worst case time while the second operation can be performed in $O(\log n)$ amortized time.
\begin{enumerate}
\item
Unit-disk range emptiness (UDRE) query: Given a point $p$, determine whether there exists a point $q \in Q$ such that $\dist(p,q) \le r$, and if yes, return such a point $q$.
\item
Deletion: delete a point $q$ from $Q$.
\end{enumerate}
\end{theorem}

To obtain our solution to the minimum bottleneck moving spanning tree problem, we first observe that the decision problem is $b(T^*) \le r$ can be solved in near-linear time by applying
Chan and Skrepetos' procedure~\cite{CS}, mentioned above, for single source shortest path in unit disk graphs, together with the semi-dynamic data structure of Wang and Zhao, summarized in Theorem~\ref{thm:WZ}, where the data structure is constructed for $Q=P(1)$. (The `dilution' of the new level is done by considering the points of the previous level, one at a time. For each such point $p$, repeatedly perform a UDRE query in the data structure and delete the returned point $q$ (if found) from the data structure, until no point is found. Finally, retain only the points of the new level that were found in one of the queries performed.) 

\matya{WRONG!!! We need a data structure that will allow us to retain all points $p_j(0)$ in the new level for which there exits a point $p_i(0)$ in the previous level, such that both $\dist(p_i(0),p_j(0)) \le r$ and $\dist(p_i(1),p_j(1)) \le r$. Maybe it is not hopeless and some sort of a semi-dynamic union/intersection tree on disks of near-linear size can be used for this purpose.}

This observation, by itself, does not lead to a reduction in the running time, since the cost of distance selection, used to produce the next candidate value $r$, is $O^*(n^{4/3})$. However, by combining it with our technique, we are able to reduce the running time to $O^*(n^{6/5})$.

\begin{theorem}
\label{thm:moving_st}
The minimum bottleneck moving spanning tree problem can be solved in $O^*(n^{6/5})$ randomized expected time.
\end{theorem}
}

\subsection{Perfect matching in disk intersection graphs} \label{sec:pm}

\begin{figure}[ht]
	\centering
	\includegraphics[scale=0.7]{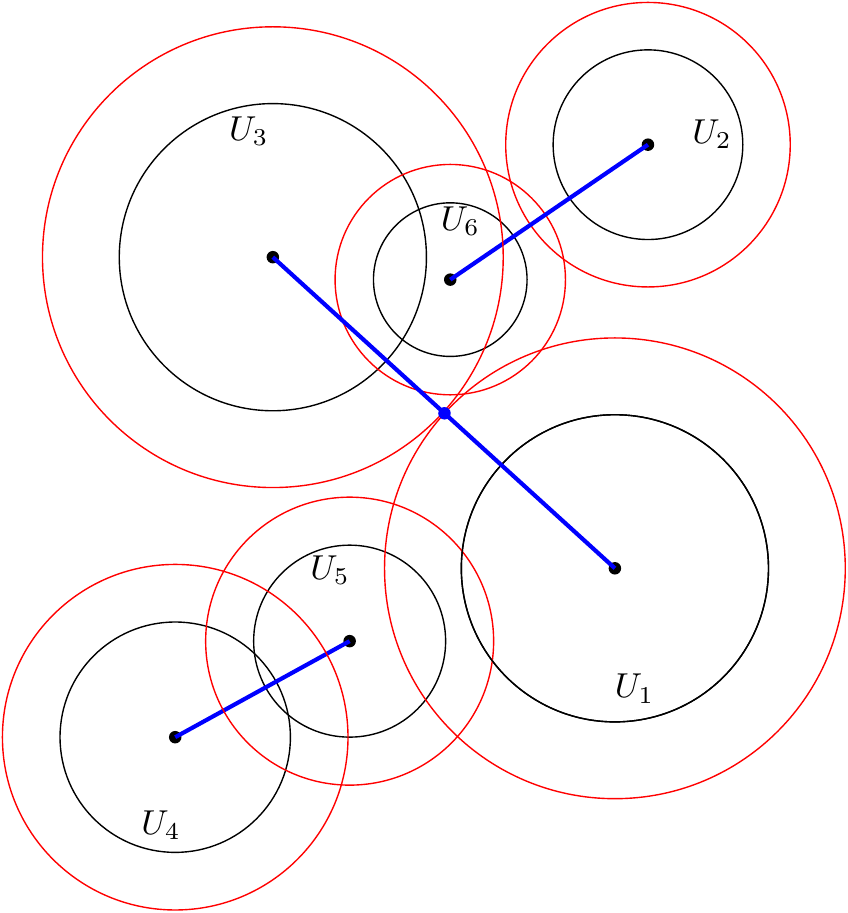}
	\caption{\sf A set $\U$ of six disks (drawn in black). In this example, $r$ is 
a multiplicative parameter, and $r^*$ is determined by the pair $U_1,U_3$. That is, if 
we blow up each of the disks of $\U$ by $r^*$, then the intersection graph of the blown 
up disks (drawn in red) has a perfect matching (drawn in blue), and $r^*$ is the smallest such value.}
	\label{fig:matching}
\end{figure}

Let $\U$ be a set of $n=2m$ disks in the plane with radii in the 
interval $[1,\Psi]$ for some constant parameter $\Psi$. The problem is to
find the smallest value $r^*$ so that if we blow up each disk about its center
by $r^*$, either additively or multiplicatively, the intersection graph of the
expanded disks has a perfect matching, see Figure~\ref{fig:matching}.

Bonnet et al.~\cite{BCM} present an algorithm for computing a matching in the 
intersection graph $G$ of $\U$ that, with high probability, is a maximum 
matching. The algorithm runs in 
\[
O(\Psi^6 n \log^{11} n + \Psi^{12\omega}n^{\omega/2}) = O(n^{\omega/2}) 
\]
expected time, where $\omega\approx 2.37286$ is the exponent of matrix multiplication.
By running the algorithm $O(\log n)$ times, we can make its failure probability
sufficiently small (polynomially small in $n$).

Let $\U_r$ denote the set of the disks of $\U$, each expanded by $r$. That is,
we either add $r$ to its radius or multiply it by $r$. We apply the procedure
of \cite{BCM} to the intersection graph $G(r)$ of $\U_r$ as a decision procedure
for the search of the optimum value $r^*$. 
We observe that the requirement that 
the radii are between 1 and $\Psi$ does not cause a problem. Indeed, when the
blow-up of the disk radii is multiplicative, the ratio between the largest and 
smallest radii does not change and remains $\Psi$. When the blow-up is additive, 
the ratio only goes down, assuming that $r^* > 0$ (that is, that the intersection
graph of the original disks does not contain a perfect matching).

\begin{figure}[ht]
	\centering
	\includegraphics[scale=0.8]{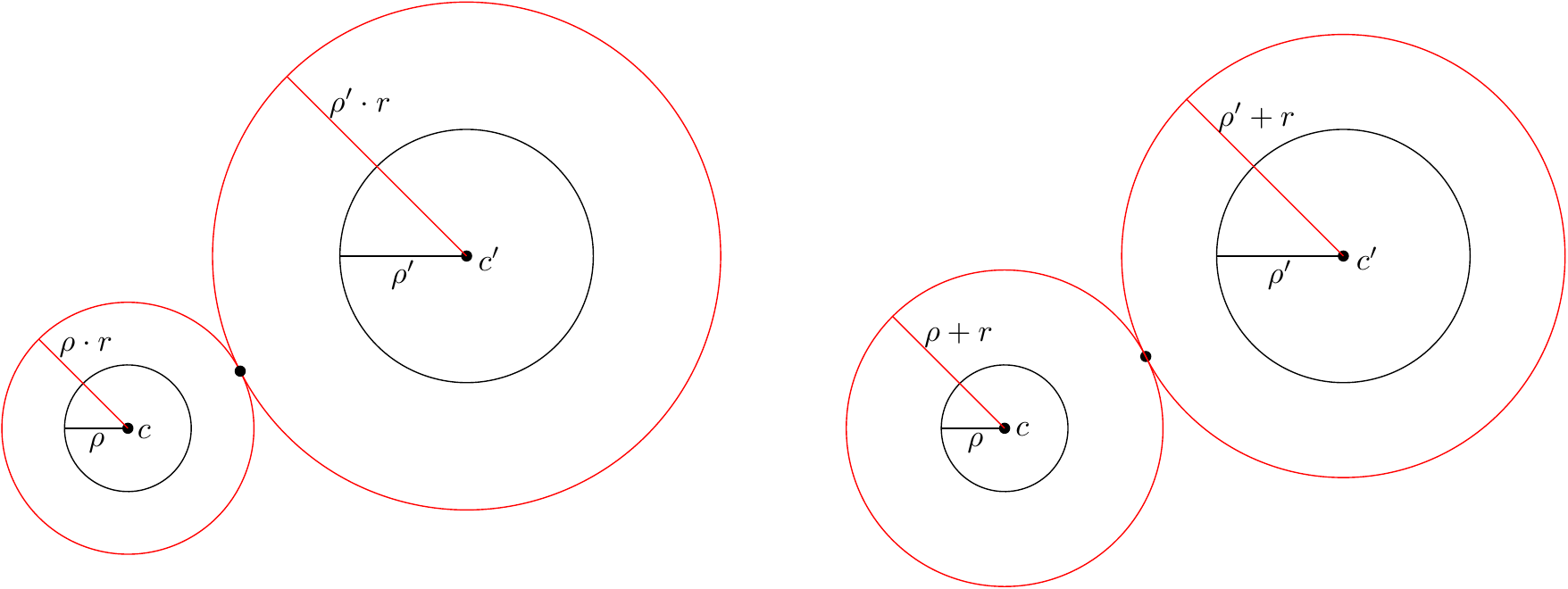}
	\caption{\sf The critical value induced by a pair of disks. 
Left: Multiplicative blow-up; $r=|cc'|/(\rho+\rho')$. Right: Additive blow-up; $r=\frac12 \left(|cc'|-\rho-\rho'\right)$.}
	\label{fig:crit_disks}
\end{figure}

The critical values here are values of $r$ at which two expanded disks touch one another, see Figure~\ref{fig:crit_disks}.
Specifically, when the blow-up is multiplicative, the critical value induced by a pair 
of disks, with respective centers $c$, $c'$ and radii $\rho$, $\rho'$, is
$|cc'|/(\rho+\rho')$. When the blow-up is additive, the critical value is
$\frac12 \left(|cc'|-\rho-\rho'\right)$. This is the first application where
the critical values are not distances.

Plugging $t=3$ (the number of degrees of freedom of a disk) and $D(n)=O^*(n^{\omega/2})$ 
in the general bound given in Theorem~\ref{th:main}, we obtain:
\begin{theorem} \label{thm:pmdisk}
In the above setting, one can compute, with arbitrarily large probability,
the smallest $r^*$ such that $G(r^*)$ contains a perfect matching, in 
$O^*(n^{3/4+\omega/4}) \approx O(n^{1.3432})$ randomized expected time. 
\end{theorem}


\medskip
\noindent{\bf Perfect matching in intersection graphs of fat planar objects of similar size.}
%
Actually, the $O^*(n^{\omega/2})$ bound mentioned above holds for any set of \emph{fat} 
objects of similar sizes, provided that one can perform some basic operations on these
objects efficiently. For example, suppose that $\U$ is a set of $n=2m$ ellipses 
in the plane, such that for each ellipse in $\U$, the ratio between its major 
and minor axes is at most $\Psi_1$, and the length of its major axis is in 
$[1,\Psi_2]$, for some constant parameters $\Psi_1, \Psi_2$. Then one can compute 
the smallest $r^*$ such that $G(r^*)$ contains a perfect matching, 
in $O^*(n^{(5+2\omega)/7}) \approx O^*(n^{1.3922})$ expected time 
(here the number of degrees of freedom for an ellipse is $t=5$ and $D(n) = O^*(n^{\omega/2})$). 
If the ratio between the major and minor axes of the ellipses is some constant $c$, 
then $t$ reduces to $4$, and the running time improves to
$O^*(n^{(8+3\omega)/11}) \approx O^*(n^{1.3744})$ randomized expected time.

\old{
\paragraph{Perfect matching in intersection graphs of balls in higher dimensions.}

\micha{Please expand.}

Let $\U$ be a set of $n=2m$ unit balls in $\reals^d$. Bonnet et al~\cite{BCM} 
present a randomized algorithm that computes a maximum matching in the intersection 
graph $G$ of these balls in $O(n^{3/2})$ time, for $d =3,4$, and in 
$O^*\left(n^{\frac{2\lceil d/2 \rceil}{1+\lceil d/2 \rceil}}\right)$ time, 
for $d \ge 5$. (The sparsification step in their algorithm, which replaces
the intersection graph by a subgraph with a linear number of edges, takes $O^*(n^{4/3})$
time for $d=3,4$, and $O^*\left(n^{\frac{2\lceil d/2 \rceil}{1+\lceil d/2 \rceil}}\right)$
time for $d\ge 5$. For $d\ge 5$, the maximum matching in the sparsified graph is 
computed by the algorithm of Micali and Vazirini \cite{MV} in $O(n^{3/2})$ time,
so the sparsification cost dominates the cost of the algorithm.)

For example, in $d=4$ dimensions, we can compute the smallest $r^*$ such that $G(r^*)$ 
contains a perfect matching (equivalently, a bottleneck matching $M$ of the set of the 
centers of the disks of $\U$) in $O^*(n^{17/11})$ expected time. 
(Here $t=4$ and $D(n)=O(n^{3/2})$.)
\matya{This is not interesting if there exists, as in the planar case, a linear-size graph 
that contains $M$ and that can be computed fast.}
For $d\ge5$, $t=d$ and $D(n)$ is determined by the running time of the sparsification, 
which is $O^*(n^{3/2})$ for $d=5,6$. 
}

\subsection{Generalized distance selection in the plane} \label{sec:sel}

\begin{figure}[ht]
	\centering
	\includegraphics[scale=0.8]{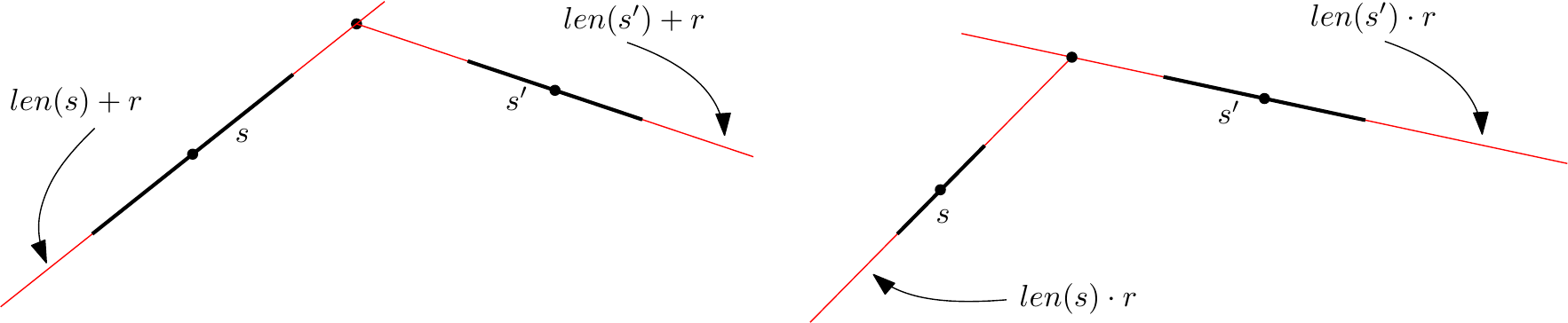}
	\caption{\sf The distance between a pair of segments $s$ and $s'$. Left: Additive expansion. Right: Multiplicative expansion.}
	\label{fig:crit_segs}
\end{figure}
 
\begin{figure}[ht]
	\centering
	\includegraphics[scale=0.8]{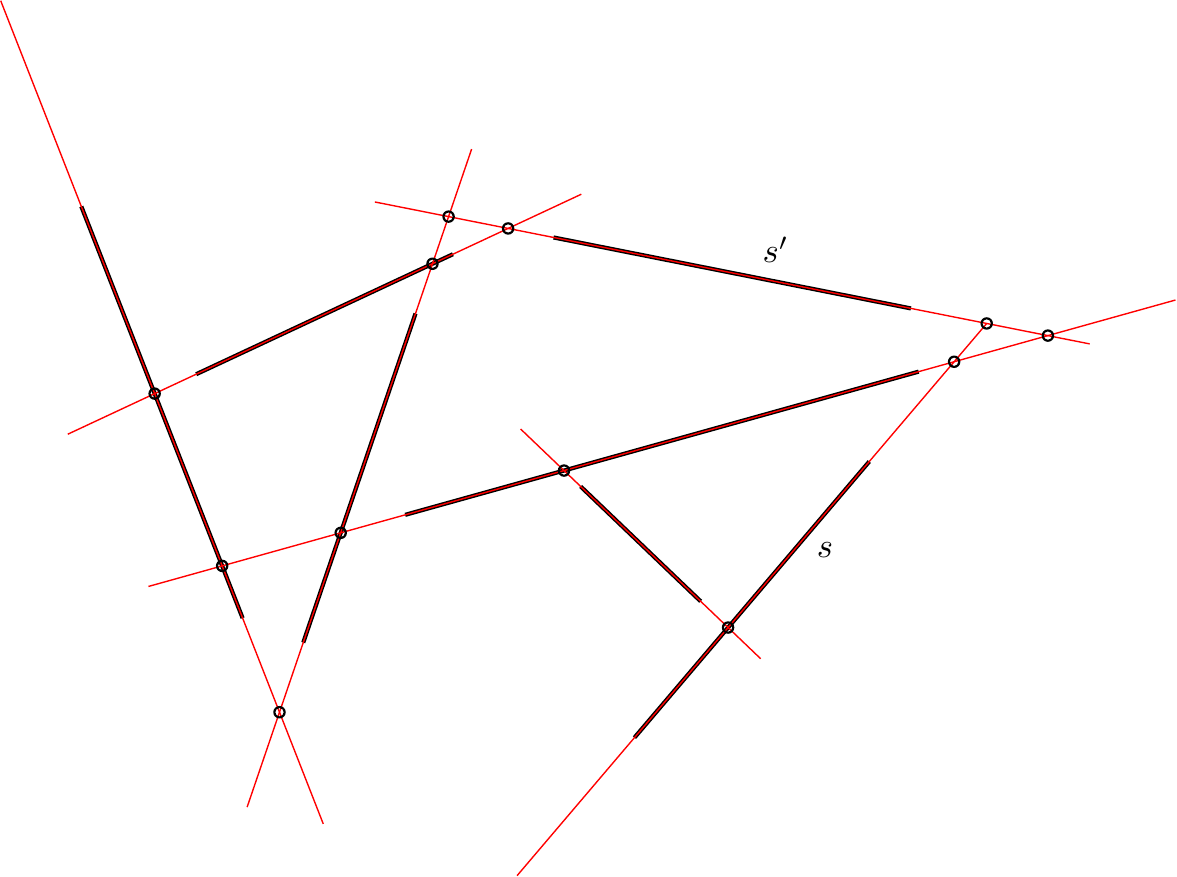}
	\caption{\sf A set $\S$ of 7 segments (drawn in black). The 12th smallest distance determined by $\S$ is the one between segments $s$ and $s'$. It is equal to the smallest $r$ for which there are exactly 12 vertices in the arrangement $\S_r$. In this example $r$ is a multiplicative parameter.}
	\label{fig:segs_arr}
\end{figure}

Let $\S$ be a set of $n$ pairwise disjoint segments in the plane. We define the distance 
between a pair of segments as the minimum value $r$ such that, if we expand the length of 
each of the segments about its center by $r$, then they intersect; see Figure~\ref{fig:crit_segs}. The expansion can be 
either additive (i.e., we add $r$ to each of the segment lengths) or multiplicative (i.e, 
we multiply each of the segment lengths by $r$). We want to find the $k$'th smallest
distance among the $n \choose 2$ pairwise distances determined by the segments in $\S$, 
for a given parameter $1 \le k \le {n \choose 2}$. This is equivalent to finding the 
smallest $r$ for which there are exactly $k$ vertices (i.e., intersection points) in the 
arrangement of $\S_r$, where $S_r$ denotes the set of the segments of $\S$, each expanded by $r$; see Figure~\ref{fig:segs_arr}.

On the one hand, this problem can be solved using the algorithm in the remark following Theorem~\ref{th:shrink}.
In this case we have $t=4$ degrees of freedom to represent a segment, which makes the algorithm run
in $O^*(n^{8/5})$ randomized expected time. 
On the other hand, the corresponding decision problem --- given $r$, is the number of vertices 
in the arrangement of $\S_r$ greater than, equal to, or smaller than $k$ --- can be solved,
using line sweeping, in $O((n+k)\log n)$ time. 
(We either perform the sweep to completion, when there are at most $k$ intersections between the
expanded segments, or stop it when the number of intersections exceeds $k$.)
Our technique allows us to combine the generalized interval shrinking procedure with 
the efficient decision procedure, and obtain a solution to the selection problem, which 
is more efficient than the aforementioned na\"ive solution when $k$ is not too large. 
More precisely, if $k = n^{1+\delta}$, we can find the $k$'th smallest distance in $\S$ 
in randomized expected time $O^*(n^{(14 + 6\delta)/11})$. Since $n^\delta = k/n$, this can be 
rewritten as $O^*(n^{8/11}k^{6/11})$, and this is faster than the na\"ive solution when
$k = o(n^{8/5})$. 

Note that the critical value determined by a pair of segments $e$, $e'$ is $\max\,\{r,r'\}$,
where $r$ is the growth parameter at which $e(r)$ touches the line supporting $e'$, and 
$e(r)$ is the segment $e$ expanded by $r$, and $r'$ is defined in a symmetric manner,
exchanging $e$ and $e'$. This is thus another instance where the critical values are not distances.

As another example, consider a set $\D$ of $n$ disjoint disks of arbitrary radii in the plane. 
We seek the smallest $r$ such that, if we expand each of the disks in $\D$ by $r$ (i.e., we 
either add $r$ to each of the radii, or multiply each of the radii by $r$), then the number 
of intersecting pairs of disks in the resulting scene is exactly $k$, for some integer 
parameter $1 \le k \le {n \choose 2}$. In other words, as in Section~\ref{sec:pm},
the distance between disks $D=D(c,r)$ and $D'=D'(c',r')$ in $\D$ is $r=(|cc'|-(r+r'))/2$ 
in the additive variant, and $r=|cc'|/(r+r')$ in the multiplicative variant. 
We seek the $k$'th smallest distance determined by the disks in $\D$. 

In this case, the problem can be solved using the same algorithm as in the previous case.
Here disks have $t=3$ degrees of freedom, so the algorithm runs in $O^*(n^{3/2})$ randomized expected time. 
The corresponding decision problem can be solved using line sweeping, as above, in $O((n+k)\log n)$ time. 
Again, our technique allows us to combine the general interval-shrinking mechanism with 
this efficient decision procedure, to obtain a more efficient solution, provided that 
$k$ is not too large. More precisely, if $k = n^{1+\delta}$, we can find the $k$'th smallest 
distance in $\D$ in $O^*(n^{(5 + 2\delta)/4}) = O^*(n^{3/4}k^{1/2})$. This is faster than
the na\"ive solution when $k = o(n^{3/2})$.

We have considered only two of the many settings in which one can obtain improved bounds 
for selecting the $k$'th smallest distance, for a sufficiently small parameter $k$, 
analogously defined for a set of pairwise-disjoint geometric objects in the plane. 
The results that we did obtain above are summarized in the following theorem.
\begin{theorem} \label{thm:distx}
(i) For $k = o(n^{8/5})$, one can compute the $k$'th smallest distance in a set $\S$ of $n$
pairwise-disjoint segments in the plane, as defined above, 
in $O^*(n^{8/11}k^{6/11})$ randomized expected time.

\noindent
(ii) For $k = o(n^{3/2})$, one can compute the $k$'th smallest distance in a set $\D$ of
$n$ pairwise-disjoint disks in the plane, as defined above, in 
$O^*(n^{3/4}k^{1/2})$ randomized expected time. 
\end{theorem}

\subsection{The discrete one-sided Fr\'echet distance with shortcuts \\ in higher dimensions} \label{app:dfds-hd}

We can extend the analysis in \cite{BFKKS} to higher-dimensional spaces $\reals^d$. 
We have two sequences $A = (a_1,\ldots,a_n)$ and $B = (b_1,\ldots,b_n)$ of points in $\reals^d$.
(We assume for simplicity that $|A|=|B|=n$.) The goal is to find a sequence of pairs
$(a_{i_1},b_{j_1}),\; (a_{i_2},b_{j_2}),\cdots, (a_{i_k},b_{j_k})$, so that
(i) ${i_1} = 1$, ${j_1} = 1$, ${i_k} = n$, and ${j_k} = n$,
(ii) both sequences $(i_1,\ldots,i_k)$, $(j_1,\ldots,j_k)$ are weakly monotone increasing, 
(iii) $i_{t+1} = i_t$ or $i_t+1$ for each $t < k$, and $\max_t \{ \dist (a_{i_t},b_{j_t} ) \}$
is minimized. The latter quantity is called the 
\emph{discrete one-sided Fr\'echet distance with shortcuts} (where the shortcuts are allowed
only for $B$), and is denoted as $\dfds(A,B)$. See \cite{BFKKS} for more details.

This is another instance of an optimization problem whose critical values are distances between
points, except that here we are concerned only with `bichromatic' distances between the points of $A$
and those of $B$. The decision procedure is, given a threshold $\eps>0$, to determine whether
$\dfds(A,B) \le \eps$. As shown in \cite{BFKKS}, this procedure can be performed in $O(n)$ time,
by simply searching for the existence of a weakly monotone path of a certain kind, from entry $(1,1)$
to entry $(n,n)$ in a zero-one $n\times n$ matrix, whose $(i,j)$-entry is $0$ (resp., $1$)
if $\dist(a_i,b_j) > \eps$ (resp., $\dist(a_i,b_j) \le \eps$). Checking for the existence of
such a path can be done in linear time, exactly as in \cite{BFKKS}. Clearly, this part takes
linear time in any dimension.

Plugging $t=d$ and $D(n)=O(n)$ in the general bound given in Theorem~\ref{th:main}, we obtain:
\begin{theorem}
The discrete one-sided Fr\'echet distance with shortcuts between two sets of $n$ points
in $\reals^d$ can be computed in $O^*(n^{(4d-2)/(3d-1)})$ randomized expected time.
\end{theorem}

\subsection{The reverse shortest path problem for unit-ball graphs \\ in three dimensions} \label{sec:sp3d}

The unweighted and weighted variants of this problem are obvious extensions to three 
dimensions of the corresponding reverse shortest path problems in the plane. That is, 
we are given a set $P$ of $n$ points in $\reals^3$, two points $s,t\in P$, and a 
parameter $1 \le k \le n-1$ (or, in the weighted variant, a length $w \ge \dist(s,t)$), 
and the goal is to compute the smallest value $r^*$ so that there exists a path between $s$ and $t$ of 
at most $k$ edges (resp., of length at most $w$) in the graph $G(r^*)$ over $P$, whose edges are
all the pairs $p,q\in P$ such that $\dist(p,q) \le r^*$.

Unlike the problem of the discrete one-sided Fr\'echet distance with shortcuts, whose 
decision procedure runs in $O(n)$ time in any fixed dimension $d$, 
the running time of the suitable extension to higher dimensions of both the decision 
procedure of \cite{CS}, for the unweighted planar reverse shortest path problem, 
and that of \cite{WangX20}, for its weighted variant, increases in $d$ as we move 
to higher dimensions. We show below that both procedures can be adapted to three 
dimensions, so that the resulting running time is $O^*(n^{4/3})$ for both variants.

Plugging this bound into the three-dimensional case in Theorem~\ref{th:main}, the overall 
running time of both algorithms is $O^*(n^{3/4}D(n)^{1/2}) = O^*(n^{17/12})$. That is, we have

\begin{theorem} \label{thm:rsp3d}
The unweighted reverse shortest path problem, as well as its weighted variant, for unit-ball graphs 
in three dimensions on an input set $P$ of $n$ points can be solved in $O^*(n^{17/12})$ randomized expected time.
\end{theorem}

To complete the presentation, we briefly describe how to adapt the planar decision procedures to three dimensions.

\medskip
\noindent{\bf The unweighted variant.}
Most steps of the procedure of \cite{CS} carry over to three dimensions with 
straightforward modifications, without affecting their asymptotic cost, except 
for the following step in the efficient BFS implementation, which in three dimensions 
reads as follows. We have a set $\S$ of $u$ congruent spheres (say for concreteness, 
of radius $1$), whose centers lie on one side of some given axis-parallel plane, say 
below the $xy$-plane, and a set $Q$ of $v$ points that lie on the other side of the 
plane (above the $xy$-plane), and the goal is to report all points of $Q$ that lie 
below the upper envelope of $\S$. 

This step can be implemented to run in $O^*(u^{2/3}v^{2/3} + u + v)$ time,
as follows. We pass to the dual space, where the points of $Q$ are mapped to unit
spheres centered at these points, and the spheres of $\S$ become points (the centers 
of these spheres). Denote the set of dual spheres as $Q^*$ and the set of dual points 
as $\S^*$. Now the goal is to determine, for each sphere in $Q^*$, whether
it contains a point of $\S^*$; this holds if and only if the primal
point of $Q$ lies below the upper envelope of $\S$.

This dual problem can be solved by lifting the spheres and points to 
$\reals^4$, so that the spheres become hyperplanes, and the problem
becomes that of testing each lower halfspace bounded by such a hyperplane
for emptiness (of the lifted points). As shown in \cite{Ma:rph}, 
this can be done in $d=4$ dimensions using $s$ storage and $O^*(s)$ 
preprocessing time, so that each emptiness query takes 
$O^*(u/s^{1/\lfloor d/2\rfloor}) = O^*(u/s^{1/2})$ time, for a 
total cost of $O^*(s + uv/s^{1/2})$. With a suitable choice of $s$
this becomes $O^*(u^{2/3}v^{2/3} + u + v)$, as claimed.

By a suitable adaptation of the technique of \cite{CS}, the overall
cost of this step of the decision procedure is $O^*(n^{4/3})$, and
this dominates its total running time $D(n)$.

\medskip
\noindent{\bf The weighted variant.}
The procedure of \cite{WangX20} (for points in the plane) is a carefully implemented 
version of Dijkstra's algorithm. It uses a grid of cell size $\Theta(r)$ and properties 
of unit disk graphs, in a clever manner, to compute a shortest-path tree from $s$ in 
near-linear time, even though the number of edges in $G(r)$ could be quadratic. 
Informally, and not very precisely, if $c$ is the next point to be processed, then 
instead of using it to update (the shortest-path information of) its neighbors, as 
in Dijkstra's algorithm, the procedure uses \emph{all} the points in $P \cap \sigma$ 
to update \emph{all} their neighbors, where $\sigma$ is the grid cell containing $c$. 

A close and more rigorous examination of this procedure, leads to the conclusion that 
it can be adapted to three dimensions, so that, similar to the planar version, its running 
time $D(n)$ is determined by the time needed to perform a given sequence of $n$ operations, 
where each operation is either an insertion of a point in $\reals^3$ with some additive 
weight, or a nearest-neighbor query with respect to the current set of weighted points. 
Alternatively, again similar to the planar version and since we are ignoring subpolynomial 
factors, the running time of the procedure in three dimensions is determined by the best 
bound for the (static) \emph{bichromatic additively-weighted nearest neighbors problem in 3-space},
in which one is given a set $R$ of $u$ red points and a set $B$ of $v$ additively-weighted 
blue points, where $u+v=n$, and the goal is to compute for each red point its additively-weighted 
nearest blue point. Agarwal et al.~\cite{AgarwalMS91} (see also \cite{AgarwalM93}) solve
this problem (for non-weighted points) in $O^*(u^{2/3}v^{2/3}+u+v)=O^*(n^{4/3})$ time. 
However, this bound also holds in the additively-weighted variant: It reduces to vertical 
ray shooting in a convex polytope defined as the intersection of $v$ halfspaces
in $\reals^5$, using a clever lifting transform from three to five dimensions described in
Aurenhammer~\cite{Aur}; see also~\cite{AgarwalM93}. This latter task, in $d=5$ dimensions,
can be solved with $s$ storage in time 
\[
O^*\left(s + \frac{uv}{s^{1/\lfloor d/2 \rfloor}}\right) = 
O^*\left(s + \frac{uv}{s^{1/2}}\right) , 
\]
which becomes $O^*(n^{4/3})$ with a suitable choice of $s$.

These arguments complete the proof of Theorem~\ref{thm:rsp3d} in both the unweighted and weighted scenarios.

\medskip
\noindent{\bf Higher dimensions.}
One can extend
the result for unweighted unit-ball graphs in any higher dimension. 
To achieve a nontrivial performance bound here, we note that the decision procedure 
can be implemented to run faster than the cost of distance selection. 
Indeed, in the
unweighted version of the problem, the decision procedure reduces to halfspace 
emptiness queries in $d+1$ dimensions, as in the three-dimensional case described 
above, whereas the distance selection reduces to range searching (with spherical 
shells) in $d$ dimensions. With $n$ objects and $s$ storage, the former task has 
query time $O^*(n/s^{1/\lfloor (d+1)/2\rfloor})$, whereas the latter task has 
query time $O^*(n/s^{1/d})$. Since $\lfloor (d+1)/2\rfloor$ is always smaller 
than $d$, the decision procedure should indeed be faster than distance selection.
Concretely, choosing the right value of $s$, the cost of the decision procedure is
$O^*\left(n^{\frac{2\lfloor (d+1)/2\rfloor}{1+\lfloor (d+1)/2\rfloor}}\right)$.
Plugging this into Theorem~\ref{th:main}, the overall algorithm runs in
\[
O^*\left(n^{2d/(3d-1)}D(n)^{(2d-2)/(3d-1)}\right) =
O^*\left(n^{\frac{(6d-4)\lfloor (d+1)/2\rfloor + 2d} {(3d-1)\left(\lfloor (d+1)/2\rfloor + 1 \right)} }\right) 
\]
randomized expected time,
which is indeed smaller than the cost $O^*(n^{2d/(d+1)})$ of the distance selection. 
It would be interesting to obtain a similar bound for the weighted case too.

\section{Seeing the most without being seen}
\label{sec:mpl}

\begin{figure}[ht]
\centering
\includegraphics[scale=0.8]{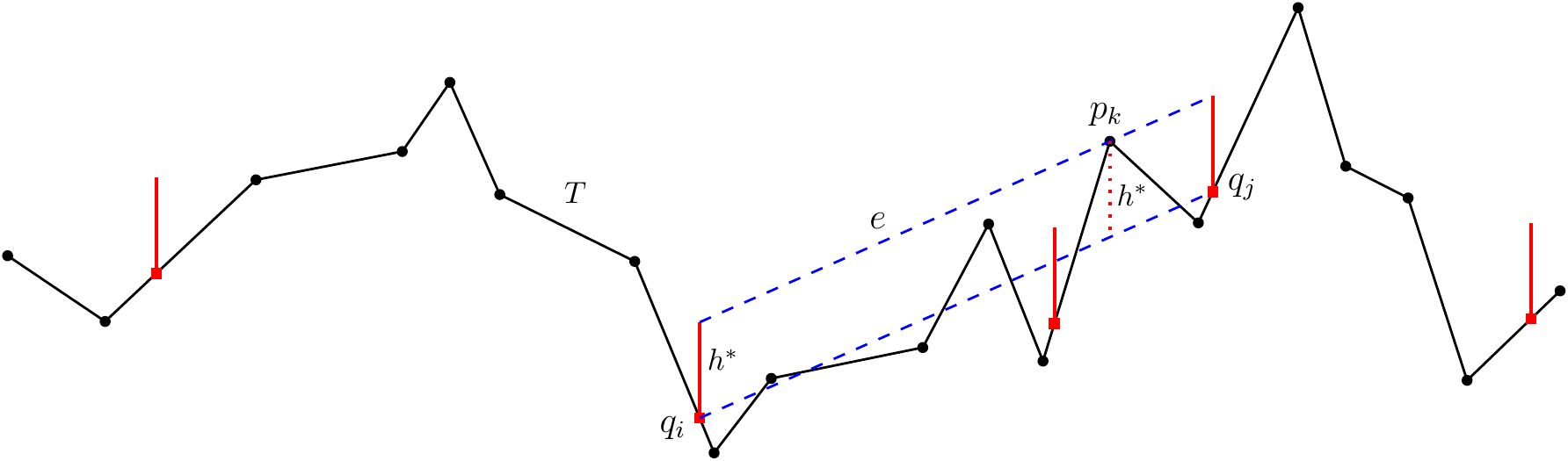}
\caption{\sf The maximum-height independent towers problem.} 
\label{fig:towers}
\end{figure}

In this section we present another application of our technique, in which the 
critical values are determined by \emph{triples} of input points, rather than by pairs.
Specifically, we study the following \emph{maximum-height independent towers} problem. 
Let $T=(p_1,\ldots,p_n)$ be a \emph{1.5-dimensional terrain}, namely a bounded 
$x$-monotone polygonal line (polyline for short). For any two points $a,b$ 
\emph{above} $T$, we say that $a$ and $b$ \emph{see each other} if the line segment 
$ab$ lies \emph{strictly} above $T$ (we do not care what happens below $T$). 
Let $Q$ be a set of $m$ points on $T$. The problem is
to compute the maximum height $h^*$, such that if we place a tower of height $h^*$ at each 
of the points $q\in Q$ (which is a vertical segment of length $h^*$ whose bottom endpoint is $q$), 
then the tips of these towers do not see each other. We assume that $h^* > 0$, which implies that 
there must be at least one vertex of $T$ between any two consecutive points in $Q$ (in their $x$-order), 
and thus $m < n$. 
See Figure~\ref{fig:towers}.

Given any set $Q$ of $m$ points \emph{above} $T$, Ben-Moshe et al.~\cite{BHKM} present an 
$O((n+m)\log m)$-time algorithm to determine whether there exist two points (from $Q$) 
that see each other. We will use this algorithm as our decision procedure. 
(This procedure essentially sweeps the scene from left to right with a vertical line,
and thus appears to be inherently sequential; we do not see any obvious way to parallelize it.)

Consider the scene in which a tower of height $h^*$ is positioned at each of the points 
$q \in Q$. Then there must exist a pair of points $q_i,q_j \in Q$, such that the segment 
$e$ between the tips of the corresponding towers passes through a vertex $p_k$ of $T$,
and none of the vertices of $T$ between $q_i$ and $q_j$ lies above $e$; see Figure~\ref{fig:towers}.
In other words, the vertical distance between $p_k$ and the segment $q_iq_j$ is $h^*$, 
and it is the maximum vertical distance between any intermediate vertex and $q_iq_j$.
This implies that $h^*$ belongs to the set $D = \{ {\rm vert}(q_i,q_j) \mid q_i,\,q_j\in Q \}$,
where ${\rm vert}(q_i,q_j)$ is the maximum vertical distance between the segment $q_iq_j$ 
and the intermediate vertices $p_k$ between $q_i$ and $q_j$ that lie above $q_iq_j$.

We use a variant of a technique due to Varadarajan~\cite{Varadarajan96} 
for the following \emph{path simplification} problem. Given $T$ as above, and a 
parameter $1 \le k \le n-1$, find an $x$-monotone polyline $T'$ with at most $k$ 
edges that best approximates $T$, where the vertices of $T'$ must form a subset of 
the vertices of $T$ and include $p_1$ and $p_n$. The deviation associated with such 
a polyline $T'$ is the maximum vertical distance between a vertex of $T$ and $T'$, 
and the goal is to minimize this deviation.

Varadarajan's solution consists of appropriate decision and optimization procedures, 
both with running time $O^*(n^{4/3})$. We first sketch a non-trivial adaptation of Varadarajan's 
optimization procedure, which, combined with the aforementioned decision procedure of 
Ben-Moshe et al.~\cite{BHKM}, yields an $O^*(n^{4/3})$ algorithm for our towers problem.
We then apply our machinery, exploiting the relative 
efficiency of the decision problem, to obtain an improved $O^*(n^{6/5})$ algorithm.

\medskip
\noindent{\bf An $O^*(n^{4/3})$ algorithm.}
Let $E_Q = \{q_iq_j \mid q_i, q_j \in Q\}$ be the set of edges of the complete graph 
over $Q$. For an edge $q_iq_j \in E_Q$ and a height $h \ge 0$, we denote the segment 
connecting the tips of the towers of height $h$ based at $q_i$ and $q_j$, by $q_iq_j(h)$, 
and set $E_Q(h) = \{q_iq_j(h) \mid q_iq_j \in E_Q\}$. We seek the largest 
value $h^*$, such that each of the edges in $E_Q(h^*)$ is a \emph{non-visibility edge}, 
in the sense that it intersects $T$. The \emph{potential} $h_{ij}$ of an edge 
$q_iq_j \in E_Q$ is the maximum height $h$ such that $q_iq_j(h)$ is a non-visibility 
edge. By definition, $h^* = \min_{i<j} h_{ij}$. 

\medskip
\noindent{\bf Constructing the non-visibility edges.}
We describe an algorithm for constructing the 
set of all edges in $E_Q$ whose potential lies in some prescribed range $I = (h_1,h_2]$. 
The algorithm partitions $T$ at its median vertex $p_{\mu}$ 
(so $\mu = \lfloor n/2\rfloor$) into a left portion $T^L$ and 
a right portion $T^R$, each consisting of at most $\lceil n/2 \rceil$ edges. 
It solves the problem (of constructing non-visibility edges $q_iq_j$ with potential 
in $I$) recursively on $T^L$ and on $T^R$, and then computes, in compact form, the 
set of all edges $q_iq_j$, with $q_i\in T^L$ and $q_j\in T^R$, whose potential is 
in $I$. The partition splits the edges of $T$ evenly between the two subproblems, 
but not necessarily the points of $Q$. We denote by $Q^L$ (resp., $Q^R$) the subset 
of points of $Q$ contained in $T^L$ (resp., in $T^R$), and put $m_L = |Q^L|$ and 
$m_R = |Q^R|$, so $m_L+m_R = m$.

\begin{figure}[ht]
\centering
\includegraphics[scale=0.9]{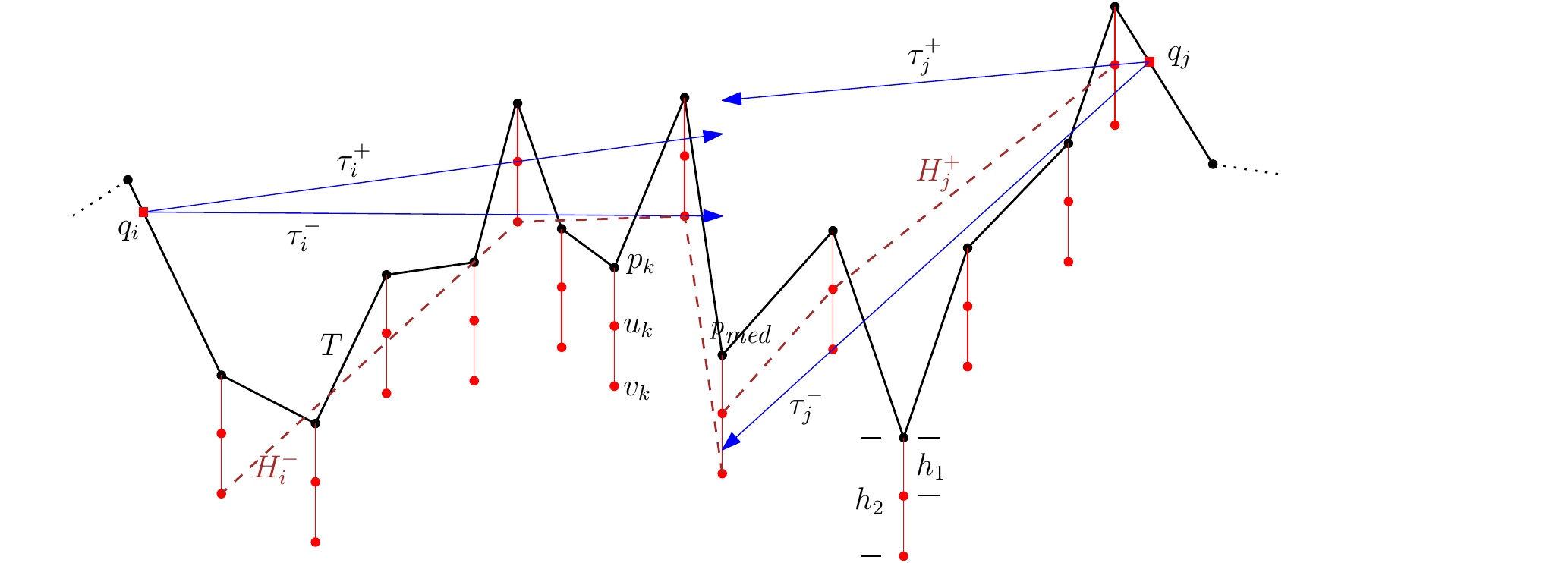}
\caption{\sf The conditions for the potential of $q_iq_j$ to be in $(h_1,h_2]$.
Here $q_iq_j$ is below $\tau_j^+$ (albeit above $\tau_i^+$) and above both 
$\tau_i^-$ and $\tau_j^-$, so the potential of $q_iq_j$ is in $(h_1,h_2]$.} 
\label{fig:hulls-and-tangs}
\end{figure}

To find the non-visibility edges between $Q^L$ and $Q^R$ we construct, for 
each vertex $p_k$, a pair of points $u_k$, $v_k$ lying on the downward 
vertical ray emanating from $p_k$ at respective distances $h_1$ and $h_2$ 
from $p_k$. For each point $q_i\in Q^L$ (resp., $q_j\in Q^R$), we consider 
the set $T^L_i = \{p_k \mid k' \le k \le \mu \}$ (resp., 
$T^R_j = \{p_k \mid \mu \le k \le k' \}$), where $k'$ is the
smallest index of a vertex of $T$ to the right of $q_i$ (resp., the
largest index of a vertex of $T$ to the left of $q_j$). Let
$H_i^-$ denote the upper convex hull of $\{v_k \mid p_k\in T^L_i\}$, and let
$H_i^+$ denote the upper convex hull of $\{u_k \mid p_k\in T^L_i\}$. 
Symmetrically, let
$H_j^-$ denote the upper convex hull of $\{v_k \mid p_k\in T^R_j\}$, and let
$H_j^+$ denote the upper convex hull of $\{u_k \mid p_k\in T^R_j\}$. 
Let $\tau_i^-$ (resp., $\tau_i^+$) denote the upper 
rightward-directed tangent ray from $q_i$ to $H_i^-$ (resp., to $H_i^+$). 
Symmetrically, let $\tau_j^-$ (resp., $\tau_j^+$) denote the upper 
leftward-directed tangent ray from $q_j$ to $H_j^-$ (resp., to $H_j^+$). 
Note that $\tau_i^-$ always lies clockwise to $\tau_i^+$,
and that $\tau_j^-$ always lies counterclockwise to $\tau_j^+$.
See Figure~\ref{fig:hulls-and-tangs} for an illustration.

It is easily seen that, by construction, the potential of $q_iq_j$ is in $I$ 
if and only if the following two conditions both hold.
\begin{description}
\item[(i)]
$q_iq_j$ passes either below $\tau^+_i$ or below $\tau^+_j$.
\item[(ii)]
$q_iq_j$ passes above both $\tau^-_i$ and $\tau^-_j$; it may overlap one of these rays.
\end{description}

\begin{figure}[ht]
\centering
\includegraphics[scale=0.8]{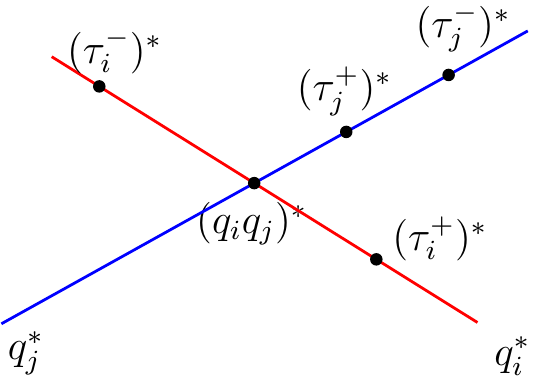}
\caption{\sf The dual setup for Conditions (i) and (ii). Condition (i) holds for $q_i$ but not for $q_j$.} 
\label{fig:dual}
\end{figure}

Passing to the dual plane, $q_iq_j$ becomes the intersection point $(q_iq_j)^*$ 
of the dual lines $q_i^*$ and $q_j^*$. The rays $\tau^-_i$ and $\tau^+_i$ are 
mapped to two respective points $(\tau^-_i)^*$ and $(\tau^+_i)^*$ on $q_i^*$, 
with $(\tau^-_i)^*$ lying to the left of $(\tau^+_i)^*$. Symmetrically,
the rays $\tau^-_j$ and $\tau^+_j$ are mapped to two respective points 
$(\tau^-_j)^*$ and $(\tau^+_j)^*$ on $q_j^*$, with $(\tau^-_j)^*$ lying 
to the right of $(\tau^+_j)^*$. Condition (i) translates to the condition 
that $(q_iq_j)^*$ lies either to the left of $(\tau_i^+)^*$ along $q_i^*$, 
or to the right of $(\tau_j^+)^*$ along $q_j^*$. Condition (ii) translates 
to the condition that $(q_iq_j)^*$ lies to the right of $(\tau_i^-)^*$ along 
$q_i^*$, and to the left of $(\tau_j^-)^*$ along $q_j^*$. See Figure~\ref{fig:dual}.

We thus face a variant of the 
classical red-blue segment intersection problem (see, e.g., \cite{Ag:rs}), in which
we have a collection $R$ of red segments $r_i := (\tau^-_i)^* (\tau^+_i)^*$, for $q_i\in Q^L$,
and a collection $B$ of blue segments $b_j := (\tau^-_j)^* (\tau^+_j)^*$, for $q_j\in Q^R$, 
and we want to collect, in compact form, all pairs $(r_i,b_j)\in R\times B$ that satisfy
Conditions (i) and (ii). As in the segment intersection problem, this can be done using
a four-level batched halfplane range searching structure, where each level enforces 
one of the four sub-conditions in (i) and (ii), each of which amounts to requiring
an endpoint of one segment (say, red) to lie in a suitable side of the line supporting
another segment (say, blue). Using standard range searching machinery (again, see \cite{Ag:rs}),
we can represent the collection of the desired pairs of segments as the disjoint union
of complete bipartite graphs $R_\alpha\times B_\alpha$, for $R_\alpha\subset R$ and 
$B_\alpha\subset B$, so that the overall size of the vertex sets of these graphs is
$O^*(m_L^{2/3}m_R^{2/3} + m_L + m_R)$. This also bounds the cost of this subprocedure.

The tangent rays $\tau^-_i$, $\tau^+_i$, for $q_i\in Q^L$, and the symmetric rays 
$\tau^-_j$, $\tau^+_j$, for $q_j\in Q^R$, from which we obtain the desired segments 
$r_i := (\tau^-_i)^* (\tau^+_i)^*$ and $b_j := (\tau^-_j)^* (\tau^+_j)^*$, are easy 
to construct in $O(n\log^2 n)$ time, as follows.

Consider the construction of the rays $\tau^-_i$; the construction of the
other three kinds of rays is done in a fully analogous fashion. Fix the index 
$i$ and put $V_i = \{v_k \mid k'\le k\le \mu\}$, where $k'$ is the index 
of the first vertex of $T^L$ to the right of $q_i$. Recall that $\tau^-_i$ is the 
line passing through $q_i$ and tangent from above to the convex hull $H_i$ of $V_i$.
We want to avoid computing all the polygons $H_i$, because this will take 
quadratic time (and storage). Instead, we build a balanced binary tree $\Xi$ 
over the vertices of $T^L$, and compute, at each node $\nu$ of $\Xi$, the convex 
hull, denoted $H_\nu$, of the vertices of $T^L$ at the leaves of the subtree of $\Xi$ 
rooted at $\nu$. This can be done in overall $O(n\log n)$ time (because each 
hull can be constructed in linear time, after a suitable initial sorting, and 
$\Xi$ has $O(\log n)$ levels). Now, for each $q_i$, we take the suffix $T^L_i$ of 
$T^L$ consisting of $p_{k'},\ldots,p_{\mu}$, and represent it as the 
disjoint union of $O(\log n)$ subtrees of $\Xi$. We compute, for each of the 
corresponding polygons $H_\nu$, its upper tangent through $q_i$, in $O(\log n)$ 
time per polygon, and set $\tau^-_i$ to be the topmost among these tangents 
(when considering their portions to the right of $q_i$).
Altogether, applying (suitably modified versions of) this procedure to all 
four kinds of tangent lines, we compute all the desired red and blue segments 
$r_i$, $b_j$, in $O(n\log^2n)$ time. 

As just mentioned, the red-blue segment interaction mechanism constructs a four-level data structure, 
each of which is a recursive batched halfplane range searching data structure. 
We use a primal-dual construction, where at each level and each recursive phase, primal or dual,
one set of segments is represented by points and the other set by halfplanes. 
Using constant-size \emph{cuttings}, each bipartite graph constructed at that 
phase corresponds to a cell $\tau$ of the cutting, and consists of all the points 
in $\tau$ and of all the halfplanes containing $\tau$.

\medskip
\noindent{\bf Shrinking the critical interval.}
However, our real goal is to obtain a shrunken interval $I$ that contains $h^*$ and 
at most $L$ other critical distances, for some suitable value of $L$. 
We solve this problem by first solving the inverse problem: 
Given an interval $I = (h_1,h_2]$, determine whether $I$ 
contains at most $L$ critical distances. Actually we 
only require that $I$ contains at most $L$ critical distances 
determined by pairs in $Q^L\times Q^R$. We then handle recursively the distances 
determined for $Q^L\times Q^L$ and $Q^R\times Q^R$, with the goal of having at 
most $L/2$ distances in $I$ for each of the two instances. Continuing recursively in this 
manner, either all the bounds
on the number of critical distances in $I$ are determined to hold, 
and then $I$ contains at most $O(L\log n)$ critical distances, or we
reach a recursive instance for which $I$ contains too many critical
distances. In this case we can shrink $I$, similar to the way it was done in
the planar case of distances. At the end of the process we obtain a shrunken interval 
that contains $h^*$ and at most $O(L\log n)$ other critical distances.

This counting procedure is implemented using the batched range searching mechanism reviewed above.
However, to gain efficiency, we run this recursion only until the size of each subproblem
is roughly $L$. The details of the algorithm are almost identical to those given in 
Section~\ref{sec:shrink}, with obvious modifications. One less trivial
modification is that the potential associated with a segment 
$q_iq_j$, namely the quantity ${\rm vert}(q_i,q_j)$ defined above,
is not readily available, and has to be computed on the fly. 

The bifurcation-tree procedure is essentially identical to the one in Section~\ref{sec:bifur},
with a few straightforward modifications.
Combining the interval-shrinking and the bifurcation-tree parts, as we did earlier, yields
the following theorem, which summarizes our result.     
\begin{theorem} \label{thm:towers}
The maximum-height independent towers problem can be solved in $O^*(n^{6/5})$ randomized expected time.
\end{theorem}

%
%
\paragraph{Acknowledgements.}
The authors thank Pankaj Agarwal for useful discussions concerning this work.

\bibliography{shrink_and_biforc}

\end{document}